\documentclass[journal,compsoc]{IEEEtran}

\usepackage{makecell}
\usepackage{balance}
\usepackage{cite}
\usepackage{amsmath,amssymb,amsfonts}
\usepackage{algorithmic}
\usepackage{graphicx}
\usepackage{textcomp}
\usepackage{indentfirst}
\usepackage{stfloats}
\usepackage{xspace}
\usepackage{multicol}
\usepackage{multirow}
\usepackage{diagbox}
\usepackage{booktabs}
\usepackage{subfigure}
\usepackage{url}
\usepackage{makecell}
\usepackage{enumitem}
\usepackage{listings}
\usepackage[most]{tcolorbox}
\usepackage{caption}
\usepackage[ruled,linesnumbered,vlined]{algorithm2e}
\usepackage{color}
\usepackage{xcolor}
\usepackage{tikz}
\usepackage{tikz-qtree}
\usepackage{pgfplots}
\usepackage{varwidth}
\usepackage{flushend}
\usetikzlibrary{patterns}
\usepackage{url}
\newcommand\etal{{\it{et al.\ }}}
\newcommand{\tool}{\text{MEGA}\xspace}
\AtBeginDocument{%
  \providecommand\BibTeX{{%
    \normalfont B\kern-0.5em{\scshape i\kern-0.25em b}\kern-0.8em\TeX}}}

\begin{document}

\title{API Usage Recommendation via Multi-View Heterogeneous Graph Representation Learning}

\author{\IEEEauthorblockN{Yujia Chen\IEEEauthorrefmark{2},
Xiaoxue Ren\IEEEauthorrefmark{3},
Cuiyun Gao\thanks{* corresponding author.}\IEEEauthorrefmark{1}\IEEEauthorrefmark{2},
Yun Peng\IEEEauthorrefmark{3},
Xin Xia\IEEEauthorrefmark{4},
and Michael R. Lyu\IEEEauthorrefmark{3}
} \\
\IEEEauthorblockA{
\IEEEauthorrefmark{2}Harbin Institute of Technology, Shenzhen, China\\
\IEEEauthorrefmark{3}The Chinese University of Hong Kong, Hong Kong, China\\
% \IEEEauthorrefmark{5}Peng Cheng Laboratory, Shenzhen, China\\
% \IEEEauthorrefmark{6}Guangdong Provincial Key Laboratory of Novel Security Intelligence Technologies, Shenzhen, China\\
\IEEEauthorrefmark{4}Software Engineering Application Technology Lab, Huawei, China}

\IEEEauthorblockA{
gaocuiyun@hit.edu.cn, yujiachen@stu.hit.edu.cn \\ \{ypeng, xiaoxueren, lyu\}@cse.cuhk.edu.hk, xin.xia@acm.org}
}

\IEEEtitleabstractindextext{%
\begin{abstract}

Developers often need to decide which APIs to use for the functions being implemented. With the ever-growing number of APIs and libraries, it becomes increasingly difficult for developers to find appropriate APIs, indicating the necessity of automatic API usage recommendation. Previous studies adopt statistical models or collaborative filtering methods to mine the implicit API usage patterns for recommendation. However, they rely on the occurrence frequencies of APIs for mining usage patterns, thus prone to fail for the low-frequency APIs. Besides, prior studies generally regard the API call interaction graph as homogeneous graph, ignoring the rich information (e.g., edge types) in the structure graph. In this work, we propose a novel method named \textbf{\tool} for improving the recommendation accuracy especially for the low-frequency APIs. Specifically, besides \textit{call interaction graph}, \tool considers another two new heterogeneous graphs: \textit{global API co-occurrence graph} enriched with the API frequency information and \textit{hierarchical structure graph} enriched with the project component information. With the three multi-view heterogeneous graphs, \tool can capture the API usage patterns more accurately. Experiments on three Java benchmark datasets demonstrate that \tool significantly outperforms the baseline models by at least 19\% with respect to the Success Rate@1 metric. Especially, for the low-frequency APIs, \tool also increases the baselines by at least 55\% regarding the Success Rate@1 score.
\end{abstract}
\begin{IEEEkeywords}
API recommendation, multi-view heterogeneous graphs, graph representation learning
% attentive network
\end{IEEEkeywords}
}
% \keywords{\textcolor{blue}{datasets, neural networks, gaze detection, text tagging}}
% \keywords{API usage recommendation, multi-view relation graphs, attentive network}

\maketitle

\IEEEpeerreviewmaketitle

\section{Introduction}\label{sec:introduction}
    In the daily software development process, developers often use the application programming interface (API) provided by some libraries to reduce development time
% and improve development efficiency 
when implementing a function. For instance, the API $BufferedInputStream.read()$ provides an efficient way to read data from an input stream and store the data
% it 
in a
% the 
buffer array. However, it is difficult for developers to be familiar with all APIs, because APIs are extensive in quantity 
% in a large number 
and rapidly evolving~\cite{DBLP:conf/wcre/HouY11, DBLP:journals/tse/YuBSM21}. In the past two decades, the number of Java Development Kit (JDK) APIs has increased more than 20 times (from 211 classes in the first version of 1996 to 4,403 classes in 2022)~\cite{DBLP:journals/sigsoft/Gvero13a}~\cite{document_cite}. 
% \yun{[any new statistics? —— done]}
%number of APIs is constantly increasing as the libraries are updated. When Java 1.0 was released in 1996, its java development kit (JDK) contained 211 classes and interfaces, while when Java 8.0 was released in 2014, they have increased to 4240 which is almost twenty times as that 18 years ago. 
Therefore, when selecting APIs, developers often refer to official technical documentation, raise questions on sites (e.g., Stack Overflow), or query on search engines (e.g., Google), etc.
%Therefore, when deciding which API to use in the working method, developers often resort to consulting the official API documentation, raising questions on sites such as Stack Overflow, making a query on search engines such as Google, and so on. 
%Obviously, these methods are time-consuming and ineffective because the official documentation is large with plenty of irrelevant contents, the quality of answers on Q\&A websites is uneven and meeting the needs of specific domains i.e., API usage query is difficult for general search engines.
Obviously, the whole process relies on developers' experience, and could be time-consuming since useful information is usually buried in massive contents~\cite{DBLP:journals/software/Robillard09, DBLP:conf/icsm/NasehiSMB12}.
%huge contents
% such methods are inefficient as the knowledge is buried in huge contents. 
% For example, XXXX.
%API usage query is difficult for general search engines.}

Regarding the issues above, previous studies~\cite{DBLP:conf/sigsoft/AcharyaXPX07, DBLP:conf/icse/BuseW12} propose to automatically recommend a list of API candidates according to previously-written code, which is demonstrated to be beneficial for improving the API searching process and facilitating software development.
%The task is formulated as ranking API candidates according to previously-written code.
%As demonstrated by previous studies~\cite{DBLP:conf/sigsoft/AcharyaXPX07, DBLP:conf/icse/BuseW12}, automatic API usage recommendation is beneficial for improving the API searching process and facilitating software development. The task is formulated as ranking API candidates according to
% the 
%previously-written code.
% Traditional methods
% \xiaoxue{Previous studies have focused on recommending API usage automatically.}They aim to recommend suitable APIs in the current programming position according to previous code information. 
% Specifically,
For example, MAPO~\cite{DBLP:conf/ecoop/ZhongXZPM09} and UP-Miner~\cite{DBLP:conf/msr/WangDZCXZ13} are based on mining frequent patterns clusters from collected projects to obtain common API usage patterns.
% from abundant projects.
PAM~\cite{DBLP:conf/sigsoft/FowkesS16} uses probabilistic modelling technique in API call sequence to mine usage patterns. FOCUS~\cite{DBLP:conf/icse/NguyenRRODP19} uses a context-aware collaborative-filtering~\cite{DBLP:conf/www/SarwarKKR01} technique to recommend APIs, relying on the similarity between methods and projects. GAPI~\cite{DBLP:conf/wcre/LingZX21} applies graph neural networks~\cite{DBLP:journals/tnn/ScarselliGTHM09} based collaborative filtering to exploit the relationship between methods and APIs. 
% challenge I : low-frequency APIs
However, these techniques focus on recommending commonly-used APIs, and tend to fail to mine the usage patterns of the low-frequency APIs. According to our analysis in Section~\ref{sec:motivation}, the low-frequency APIs occupy a significant proportion of all APIs. According to the statistics, the rarely-appeared APIs account for 76\% of the whole APIs in the $SH_L$ dataset~\cite{DBLP:conf/icse/NguyenRRODP19}.
% Contrary to the occupying ratio, 
Nevertheless,
% by observing recommendation results in $SH_L$ dataset, we find that 
the recommendation success rate of rare
% these 
APIs (7.9\%) is much lower than that of common APIs (54.2\%).
% \yun{[any example about the results?]}
% \yun{[existing approaches perform poorly?]}
% low called-frequency APIs 
% which occupy a significant proportion among all the APIs. As shown in Section~\ref{sec:motivation}, the rarely-appeared APIs can account for 76\% of the whole APIs in the $SH_L$ dataset. 
% Thus, the usage pattern mining of these APIs is essential for accurate recommendation. 
Thus, \textbf{how to effectively
% efficiently 
learn low-frequency APIs usage patterns
% with limited call interactions 
is a great yet under-explored challenge}~\cite{DBLP:journals/corr/abs-2112-12653}.
% % challenge II : 
Besides, the existing techniques highly rely on the homogeneous interaction information between APIs and methods,
% API-method interaction, 
ignoring the rich contextual information in source code (e.g., co-occurring APIs and hierarchical structure in
% of 
projects and packages).
% \yujia{Specifically, API pairs that appear more frequently in different methods are more likely to be used together. For example, if there is \textit{file.open()} in an API call sequence, then \textit{file.close()} is very likely to be called in following coding. Besides, methods with similar project structures have more consistent API calls; Likewise, APIs under the same package or class are used together more frequently.}
In fact, APIs under the same package are more likely to be called together (e.g., \textit{file.open()} and \textit{file.close()}) are under the package \textit{java.io}), which is important external information for API recommendation.
Therefore, \textbf{how to involve contextual information in API recommendation is another challenge}.
% \xiaoxue{\it{[HERE NEED to add some content to illustrate the importance of adding richer information to demonstrate the second challenge... ]}}
% Hence, how to capture abundant heterogeneous information to provide more accurate recommendation is also challenging.
% \yun{[The challenge of accurately recommending low-frequency APIs?]}, which is a great yet \yun{under-explored}
% unexplored  
% challenge~\cite{DBLP:journals/corr/abs-2112-12653}. 
% \yun{Besides, [another issue of previous studies as mentioned in the abstract?]}
% make up the majority of the APIs. 
% They tend to recommend common frequently used APIs, while ignoring the current specific functional requirements in working method, which eventually affects the effectiveness of the recommendation. See the motivating details in Section~\ref{sec:motivation}. 

% \yujia{To address the above two challenges,}

In this work, we propose \tool, a novel API usage recommendation method with \textbf{M}ulti-view h\textbf{E}terogeneous \textbf{G}raph represent\textbf{A}tion learning. Different from the prior studies, \tool employs heterogeneous graphs, which are constructed from multiple views, i.e., method-API interaction from local view, API-API co-occurrence from global view, and project structure from external view. Specifically, \tool builds upon three heterogeneous graphs, i.e., the common \textit{call interaction graph}, and two new graphs, i.e., \textit{global API co-occurrence graph} and \textit{hierarchical structure graph}. The \textit{call interaction graph} establishes the relations between methods and corresponding called APIs, and is commonly adopted by previous approaches~\cite{DBLP:conf/icse/NguyenN15, DBLP:conf/kbse/Gu0019, DBLP:conf/sigsoft/FowkesS16, DBLP:conf/icse/NguyenRRODP19, DBLP:conf/wcre/LingZX21}. Models based on only such graph cannot well capture the representations of the APIs with rare called frequencies. To improve the API representations, the \textit{global API co-occurrence graph} is introduced to build the relations between APIs with the co-occurrence frequencies incorporated. To enrich the representations of APIs and methods with contextual structure information, \tool also involves the called information by projects and packages, composing the \textit{hierarchical structure graph}.
%[add reason about why considering structure graph?]
% \yujia{The project structure is important to model the context of methods and APIs. For example, methods with similar project structures have more consistent API calls; Likewise, APIs under the same package or class are used together more frequently}. Thus, \tool involves the called information by projects and packages, composing the \textit{hierarchical structure graph}.
% To mitigate the issue, we propose \tool, a novel API usage recommendation method with \textbf{M}ulti-vi\textbf{E}w \textbf{R}elat\textbf{I}on graph\textbf{S}. 
% To \yun{better learn the}
% % achieve a better 
% semantic match between client methods and APIs, 
% \tool first constructs three views of relational heterogeneous graphs, including \textit{call interaction graph}, \textit{global API co-occurrence graph}, and \textit{hierarchical structure graph}. 
A graph representation model is then proposed to learn the matching scores between methods and APIs based on the multi-view graphs.
% they are utilized as input to train a graph representation model. 
% Specifically, the \textit{call interaction graph} builds the relations between methods and corresponding called APIs, which is commonly adopted by previous approaches~\cite{DBLP:conf/icse/NguyenN15, DBLP:conf/kbse/Gu0019, DBLP:conf/sigsoft/FowkesS16, DBLP:conf/icse/NguyenRRODP19, DBLP:conf/wcre/LingZX21} for capturing the intent of methods and usage scenario of APIs. 
% The interaction information on this graph reflects original function demand of client methods and original usage scenario of APIs, which guides potential information mining in next graphs. 
% The \textit{global API co-occurrence graph} builds the relations between APIs with the co-occurrence frequencies incorporated.
% based on the co-occurrence frequencies. The co-occurrence information reflects the potential function demand of client methods and potential usage scenarios of APIs. 
% The \textit{hierarchical structure graph} builds the relations between entities in projects and packages, which considers the information of edge types. The last two graphs are involved to enrich the representations of APIs and methods learnt from the first view of graph, especially the representations of APIs with low occurrence frequencies. 
To integrate the multi-view knowledge, a frequency-aware attentive network and a structure-aware attentive network are proposed to encode the co-occurrence information and hierarchical structure, respectively.

We evaluate the effectiveness of \tool on three Java benchmark datasets consisting of 610 Java projects from GitHub and 868 JAR archives from the Maven Central Repository. In addition, we also simulate the real development scenario~\cite{DBLP:conf/icse/NguyenRRODP19} where a developer has already called some APIs in a method. Then \tool recommends APIs based on the called APIs by client methods, and calculates the evaluation metrics. The experimental results show that \tool outperforms the baseline approaches (PAM~\cite{DBLP:conf/sigsoft/FowkesS16}, FOCUS~\cite{DBLP:conf/icse/NguyenRRODP19} and GAPI~\cite{DBLP:conf/wcre/LingZX21}) by at least 19\%
% 19.85\% 
with respect to the Success Rate@1 metric. For the low-frequency APIs, \tool also achieves an increased rate at more than 55\% compared to the baselines.
% a Success Rate@1 score increased by 55.55\% at least.

In summary, our main contributions in this paper are as follows:
\begin{itemize}
%\item \xiaoxue{We are the first work focusing on ???}
\item To the best of our knowledge, we are the first work to construct multi-view
% relational 
heterogeneous graphs for more accurate API usage recommendation.
\item We propose a novel API recommendation approach named \tool, which designs a graph representation model with a frequency-aware attentive network and a structure-aware attentive network to generate enhanced representations of methods and APIs.
% , which explore invocation intention of client methods and usage feature of APIs from multi-view relation graphs.
\item We perform experiments on three benchmark datasets, and the results demonstrate that \tool outperforms the state-of-the-art API usage recommendation approaches, even for the low-frequency APIs.
% and also improves the recommendation performance of the APIs with low occurrence frequencies.

\end{itemize}
\textbf{Outline.} The rest of paper is organized as follows: Section~\ref{sec:motivation} introduces details of our motivation. Section~\ref{sec:methodology} presents the overall workflow of \tool and architecture of the graph representation model in \tool. Section~\ref{sec:setup} and Section~\ref{sec:results} are the settings and results of evaluation, respectively. Section~\ref{sec:discussion} analyzes some implications and threats to validity. Section~\ref{sec:related} succinctly describes related works. In the end, in Section~\ref{sec:conclusion}, we conclude the whole work.

\section{Motivation}\label{sec:motivation}
    % motivaton: 为什么要关注在low fre API上？
% 通过对三个数据集上的统计，我们发现
% The main criticism of this paper can be: (1) Why combining these three graphs can capture the essence of API usage recommendation? (2) How do you combine them effectively? (3) What technical contribution is there when you combine them? For (1), in your motivation, clearly state the main factors in API usage recommendation, which can later be linked with these three graphs.  For (2), elaborate Figure 3 with more insights about how the three graphs are effectively combined.  For (3), although deep learning is directly employed, interpretability of how particular API usage recommendation is trigged due to activation of which graph can perhaps be elaborated.
% \begin{figure*}[h]
%   \centering
%   \includegraphics[trim={0 1cm 2cm 0},clip,scale=0.38,width=\textwidth]{figures/motiv-fre.pdf}
%   \caption{xxx}
%   \Description{model}
% \end{figure*}

% \subsection{Motivation}
\begin{figure}[h]
  \centering
  \includegraphics[width=\linewidth]{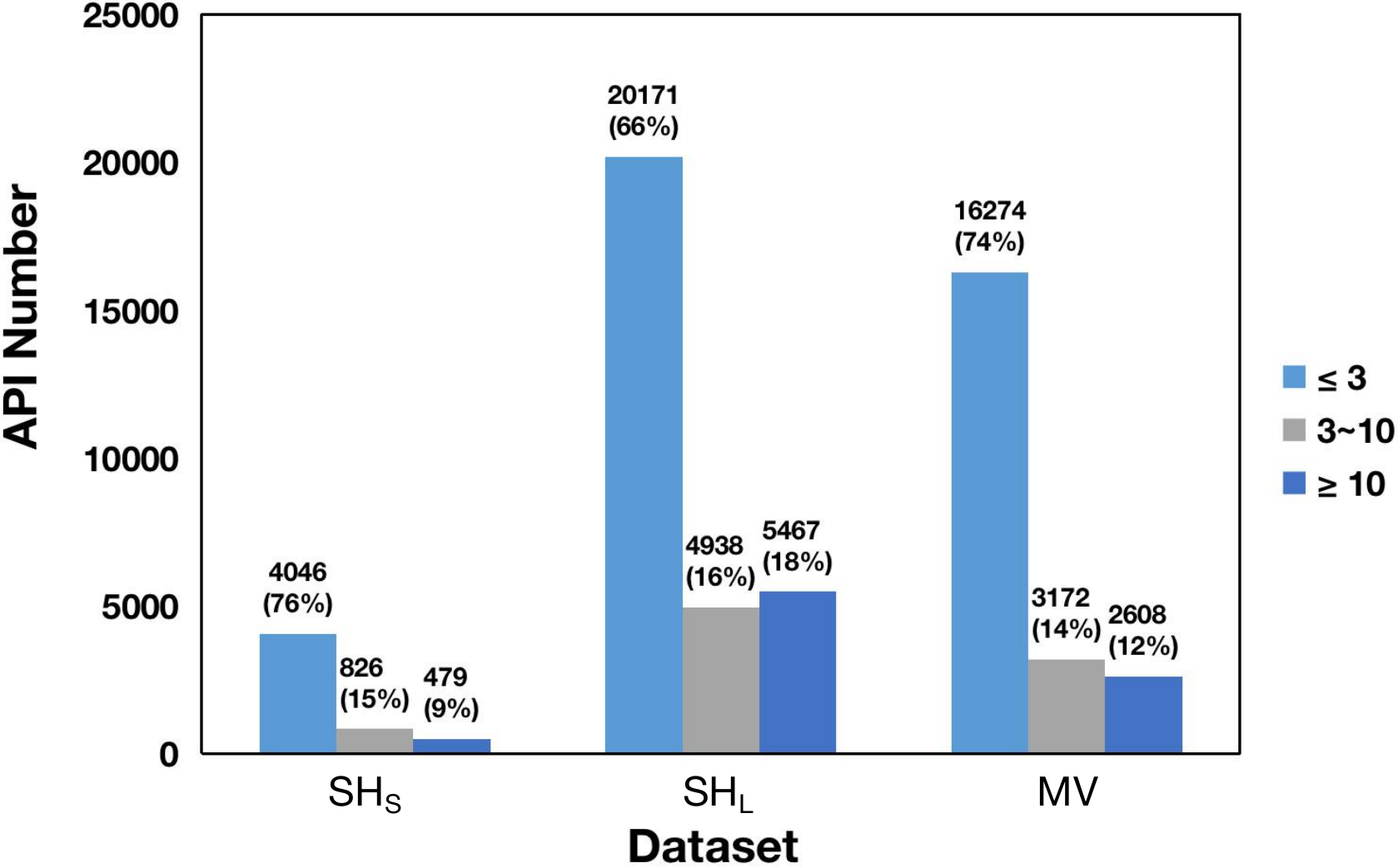}
  \caption{The distribution of APIs called frequency in $\mathbf{SH_S}$, $\mathbf{SH_L}$ and $\mathbf{MV}$ datasets.}
  \label{fig:API_call_freq}
\end{figure}

\begin{figure}[h]
  \centering
  \includegraphics[width=\linewidth]{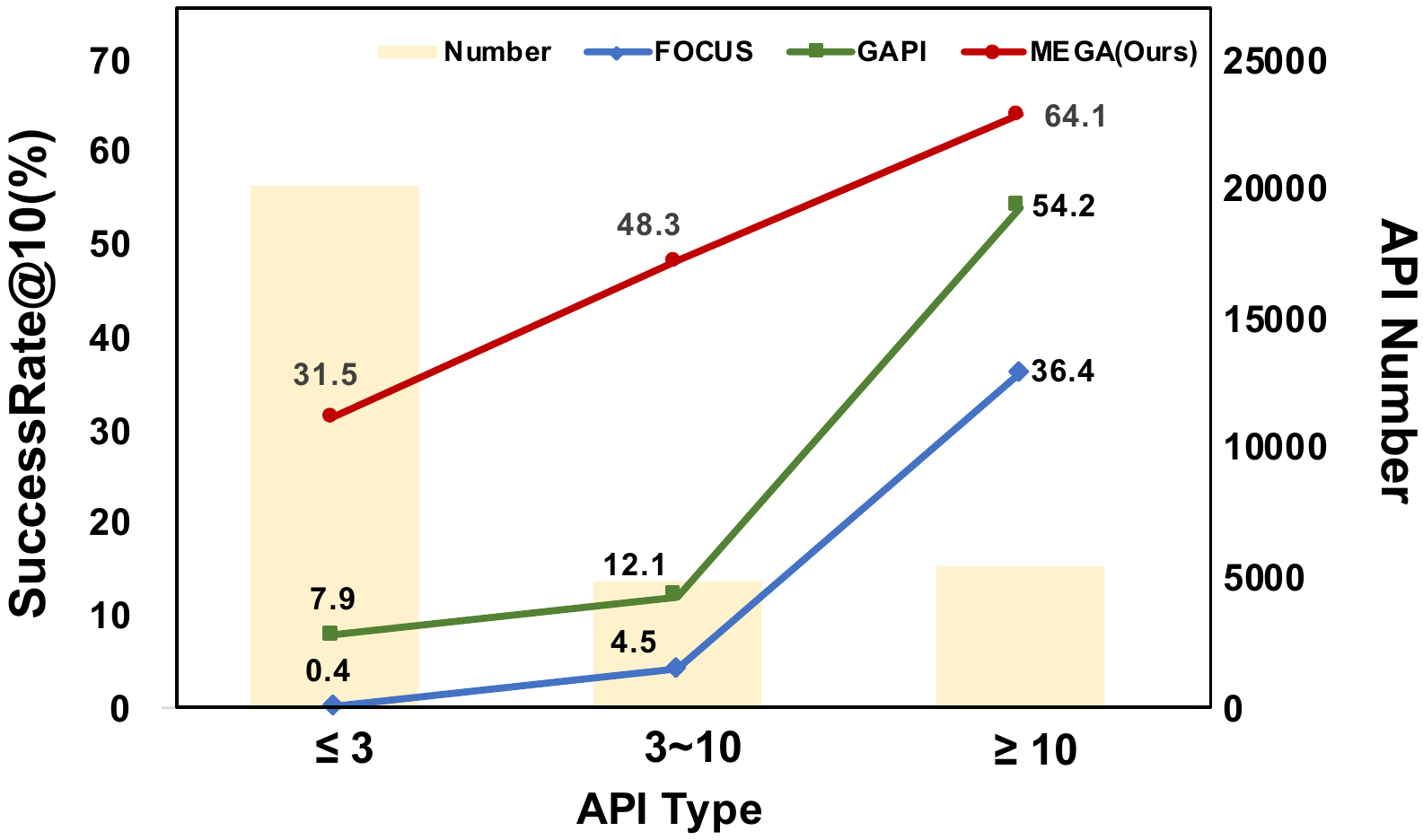}
  \caption{API recommendation performance of  FOCUS~\cite{DBLP:conf/icse/NguyenRRODP19}, GAPI~\cite{DBLP:conf/wcre/LingZX21} and our proposed \tool (corresponding to the APIs with different frequencies on the  $\mathbf{SH_L}$ dataset).}
  \label{fig:API_SR@10}
\end{figure}

% \begin{figure*}[h]
%   \centering
%   \includegraphics[trim={1cm 1cm 1cm 1cm},clip,scale=0.7]{figures/motiv-fre.pdf}
%   \caption{xxx}
%   \Description{the distribution of APIs called frequency in (a)SH\_S (b)SH\_L (c)MV}
% \end{figure*}

% \begin{figure}[h]
%   \centering
%   \includegraphics[width=\linewidth]{figures/motiv-case2.pdf}
%   \caption{xxx \xiaoxue{need a brief introduction to FOCUS and GAPI}}
%   \Description{model}
%   \label{fig:motiv_case}
% \end{figure}

% \begin{figure*}[h]
%   \centering
%   \includegraphics[trim={2cm 2cm 2cm 1cm},clip,scale=0.5]{figures/motiv-case.pdf}
%   \caption{xxx}
%   \Description{model}
%   \label{fig:motiv_case}
% \end{figure*}

Figure~\ref{fig:API_call_freq} shows the distribution of APIs with different occurrence frequencies in three benchmark
% popular 
datasets~\cite{DBLP:conf/icse/NguyenRRODP19}, i.e., $SH_S$, $SH_L$ and $MV$, with detailed statistics of the datasets shown in Table~\ref{tab:dataDes}. %in Section~\ref{sec:setup}.
Obviously, APIs with lower occurrence frequencies (i.e., $\leq 3$)
% fewer than or equal to three occurrence frequencies
account for significant proportions (i.e., $> 65\%$) among all the APIs in each dataset.
% all 
%dataset.
% interactions are the majority, accounting for more than 65\% in all datasets.
% Thus, we roughly define low occurrence frequencies as fewer than three. 
Although appearing less frequently, the large proportion of such APIs indicates developers' strong demands for specific functions, and accurately recommending the APIs is critical for facilitating their daily programming.

\begin{figure}
  \includegraphics[scale=0.425]{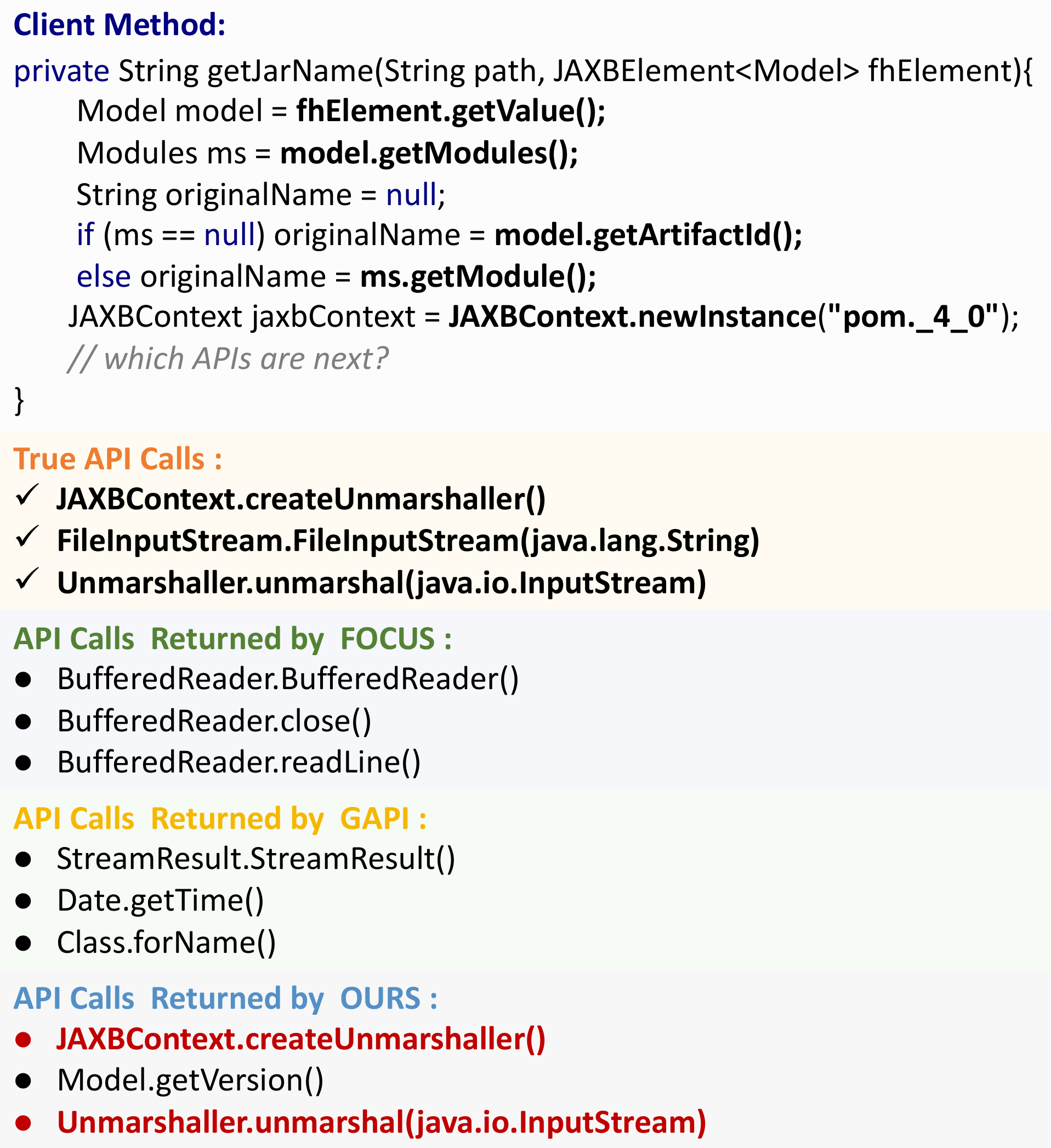}
  \caption{An Example of API usage recommendation. \emph{(The true API calls and TOP-3 APIs recommended by FOCUS~\cite{DBLP:conf/icse/NguyenRRODP19}, GAPI~\cite{DBLP:conf/wcre/LingZX21} and our proposed \tool, respectively.)}} % \xiaoxue{\it{[The first line should be Client Method]}}}
  \label{fig:example}
\end{figure}

% For instance, 
Figure~\ref{fig:example} illustrates an example of
% shows  
a client method which queries the low-frequency APIs.
% about a query recommending [\textcolor{red}{APIs with low-call frequencies}]. 
In this scenario, the developer is working on
% declares 
a method to get the name of a JAR package, but is not sure which APIs to use next.
% has partially completed the functionality, but is not sure how to proceed next.
The ``true API calls'' in Figure~\ref{fig:example} list the APIs in ground truth, in which
% real APIs developers need and we find 
both ``\textit{JAXBContext.createUnmarshalle()}'' and ``\textit{Unmarshaller.unmarshal(java.io.InputStream)}'' are rarely appear in the datasets. Both 
% the existing 
popular models including FOCUS~\cite{DBLP:conf/icse/NguyenRRODP19} and GAPI~\cite{DBLP:conf/wcre/LingZX21} learn the API representations ineffectively, and fail to recommend the APIs. Figure~\ref{fig:API_SR@10} depicts the API recommendation performance of the two models corresponding to APIs with different frequencies on the $SH_L$ dataset with respect to the SuccessRate@10 score. We find that the APIs appearing rarely, e.g., $\leq 3$, present significantly poor performance than the APIs appearing frequently, e.g., $\geq 10$. The results show that the existing models are difficult to recommend the low-frequency APIs.

Besides, existing approaches~\cite{DBLP:conf/msr/WangDZCXZ13,DBLP:journals/jcst/LingZLX19,DBLP:conf/kbse/Gu0019,DBLP:conf/icse/NguyenRRODP19,DBLP:conf/wcre/LingZX21} generally regard the API call interaction graph as homogeneous graph, ignoring the rich heterogeneous information (e.g., edge types) in the graph. For example, the state-of-the-art models, FOCUS~\cite{DBLP:conf/icse/NguyenRRODP19} and GAPI~\cite{DBLP:conf/wcre/LingZX21} are based on collaborative filtering for measuring the similarities between all methods to recommend APIs. The learning process in the models tends to rely on the commonly-used APIs in methods, resulting in ineffective API recommendation. As the example shown in Figure~\ref{fig:example}, the API recommended by FOCUS for improving the speed and efficiency of byte stream operations comes from the \textit{BufferedReader} class, which is a very general yet function-irrelevant operation for the current client method.
% limited by the single information with homogeneous graphs, it is difficult for existing approaches to recommend accurate APIs. For example, the state-of-the-art model, FOCUS and GAPI are based on collaborative filtering, measuring the similarity between all methods to recommend APIs. However, the functionality of each method is different, and thus the similarity mainly comes from the commonly-used APIs invoked in methods. As illustrated in Figure~\ref{fig:example}, FOCUS recommends APIs for improving the speed and efficiency of byte stream operations coming from \textit{BufferedReader} class, which is a very general operation in programming.

\textbf{Our approach.} To address the above limitations of the existing models,
% methods, 
we try to exploit the rich heterogeneous information in source code from multiple views, including method-API interaction from local view, API-API co-occurrence from global view, and project structure from external view, respectively. Specifically, we build three heterogeneous graphs from each view, i.e., call interaction graph, global API co-occurrence graph and hierarchical structure graph.
% such as the co-occurrence relationship of the same API pair in different methods, and the contextual package and project structure of APIs and methods. 
Moreover, two new attentive networks are designed for encoding frequency-based co-occurrence information and structure-based hierarchical information during learning the representations of APIs and methods. As the example shown in Figure~\ref{fig:example}, \tool captures the co-occurring pattern with the API ``\textit{JAXBContext.NewInstance(\"POM.\_4\_0\")}'' and their similar structural information (i.e., under the same class), so it successfully recommends the rare API ``\textit{JAXBContext.CreateUnmarshaller()}''.

\section{Methodology}\label{sec:methodology}
    
\begin{figure}
  \centering
  \includegraphics[width=0.5\textwidth]{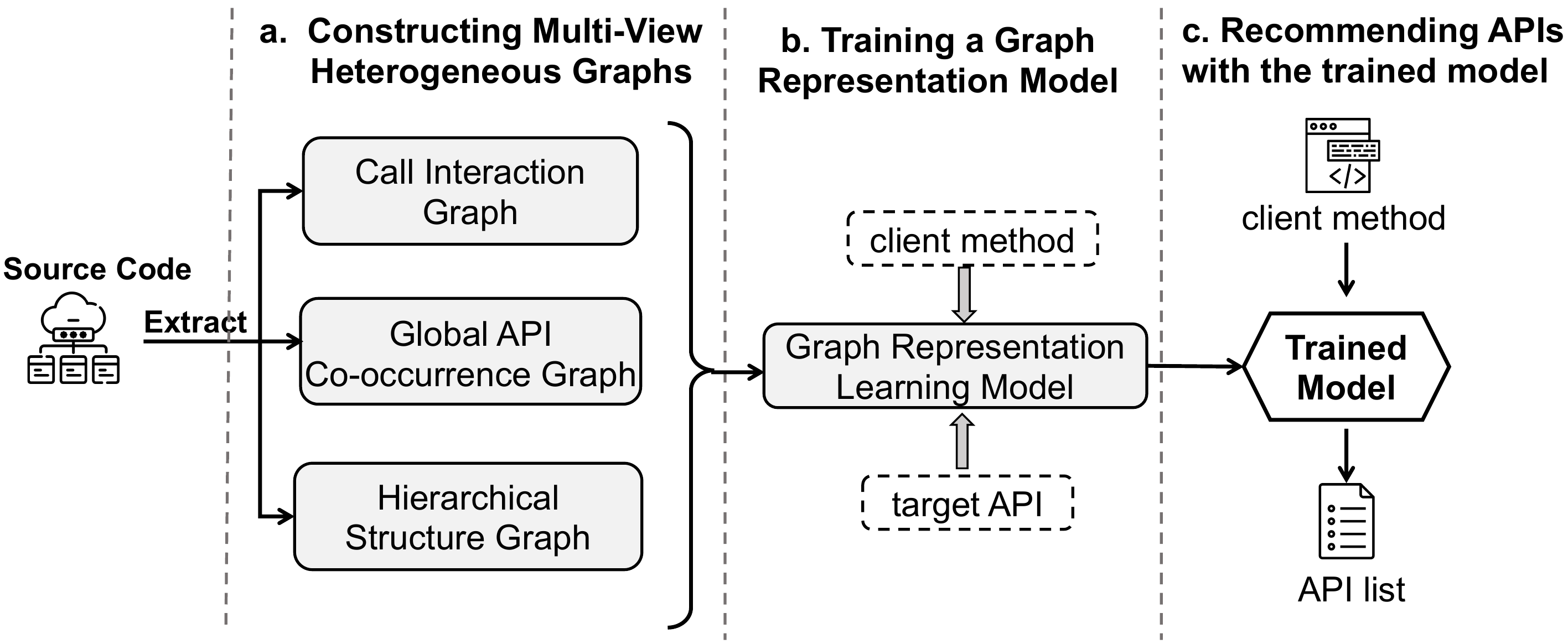}
  \caption{The overall workflow of \tool.}
  \label{fig:workflow}
\end{figure}

% \begin{figure*}
%   \centering
%   \includegraphics[scale=0.4]{figures/method-OverallWorkflow-1.pdf}
%   \caption{The Overall Workflow of \tool}
%   \label{fig:workflow}
%   \Description{model}
% \end{figure*}

% \begin{figure*}[h]
%   \centering
%   \includegraphics[scale=0.42,width=\textwidth]{figures/model.pdf}
%   \caption{Architecture of the graph representation model in \tool. [need detailed des.]}
%   \Description{model}
%   \label{fig:model_framework}
% \end{figure*}

%\begin{figure*}[h]
%    \subfigure[The architecture of graph representation model in %\tool.]{
%    \label{fig:model_arc}
%    \includegraphics[scale=0.35]{figures/figure_model.pdf}} 
%    \quad
%    \subfigure[The attentive networks in the model.]{
%    \label{fig:model_network}
%    \includegraphics[scale=0.35]{figures/network.pdf}} 
%  \caption{Illustration of the proposed graph representation model in \tool. The left subfigure (a) shows the model architecture, and the right subfigure (b) presents the two attentive networks in the model.}
%  \Description{model}
%  \label{fig:model_framework}
%\end{figure*}

\begin{figure*}
  \includegraphics[scale=0.4,width=\textwidth]{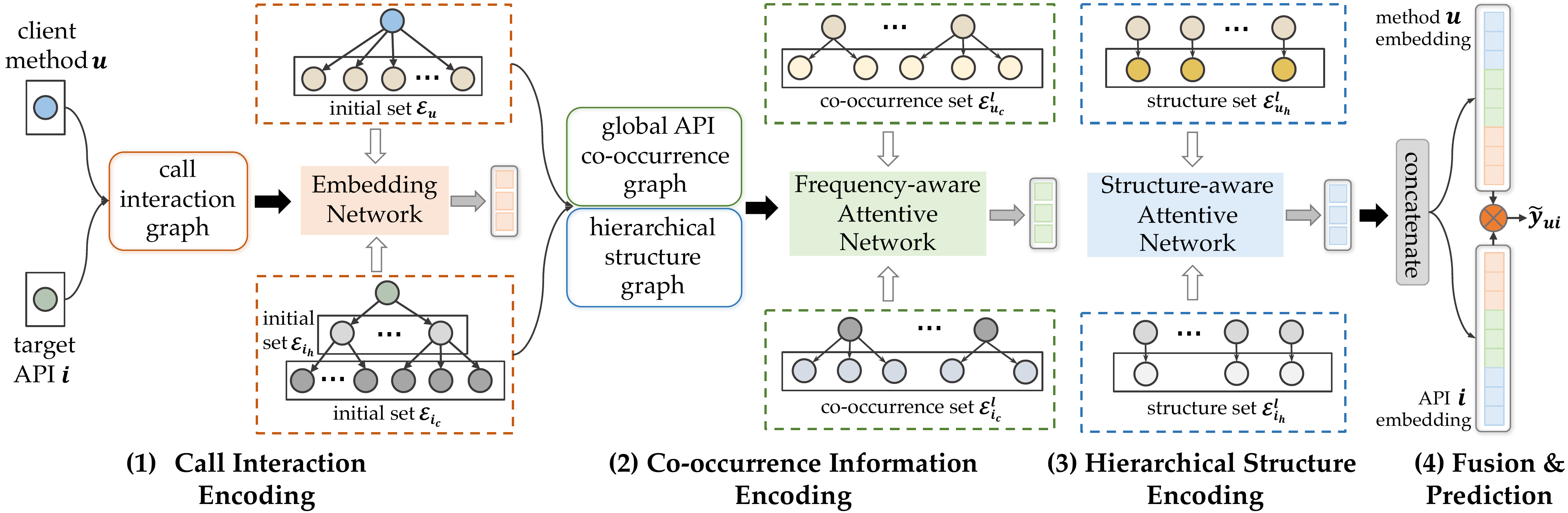}
%   \caption{Illustration of the proposed graph representation model in \tool. The left subfigure (a) shows the model architecture, and the right subfigure (b) presents the two attentive networks in the model. The red arrows mean that the initial seeds of co-occurrence information graph and hierarchical structure graph are from historical interaction.}
  \caption{Training and recommending process of the graph representation learning model in \tool.}
  %\xiaoxue{\it{[NEED RENAMING! and are the processes of Co-occurrence  and Hierarchical Structure encoding parallel?]}}}
  \label{fig:model_framework}
  
\end{figure*}

% \subfigure[The effect of hop numbers]{
% \label{fig:res-para-h}
% \includegraphics[scale=0.23]{figures/layer.pdf}} 
% \quad
% \subfigure[The effect of bucket numbers]{
% \label{fig:res-para-b}
% \includegraphics[scale=0.23]{figures/bucket.pdf}} 

\begin{table}
  \centering
  \normalsize
  \small
  \caption{Symbols and corresponding descriptions.}
  \label{tab:symDes}
  \begin{tabular}{c|c}
    \Xhline{1pt}
     \textbf{Notation} & \textbf{Description} \\
    \Xhline{0.5pt}
    $\mathcal{U}$,$\mathcal{I}$ & The set of methods, APIs \\
    \Xhline{0.5pt}
    $\mathcal{E}$,$\mathcal{R}$ & The set of entities, relations \\
    \Xhline{0.5pt}
    $\mathcal{G}_I$ & The call interaction graph \\
    \Xhline{0.5pt}
    $\mathcal{G}_C$ & The global API co-occurrence graph \\
    \Xhline{0.5pt}
    $\mathcal{G}_H$ & The hierarchical structure Graph \\
    \Xhline{0.5pt}
    $(i,f,j)$ & The co-occurrence triple \\
    \Xhline{0.5pt}
    $(h,r,t)$ & The structure triple \\
    \Xhline{0.5pt}
    $u,i$ & The client method, the target API \\
    \Xhline{0.5pt}
    $L$ & The hop number of information encoding \\
    \Xhline{0.5pt}
    $\mathcal{E}_u^l$ & The $l$-hop entity set of $u$ \\
    \Xhline{0.5pt}
    $\mathcal{S}_u^l$ & The $l$-hop triple set of $u$ \\
    \Xhline{0.5pt}
    $e_u^{(o)},e_i^{(o)}$ & The local-view representation of the method $u$,$i$ \\
    \Xhline{0.5pt}
    $e_u^{(c)},e_i^{(c)}$ & The global-view representation of $u$,$i$ \\
    \Xhline{0.5pt}
    $e_u^{(h)},e_i^{(h)}$ & The external-view representation of $u$,$i$ \\
    \Xhline{0.5pt}
    $e_u,e_i$ & The fianl representation of $u$,$i$  \\
    \Xhline{1pt}
 \end{tabular}
\end{table}

\subsection{Workflow of \tool}
Figure~\ref{fig:workflow} illustrates the \tool's workflow which includes three main stages, i.e.,
% In this section, we introduce the three main stages in \tool's workflow as shown in Figure~\ref{fig:workflow}, including 
\textbf{constructing multi-view heterogeneous graphs}, \textbf{training a graph representation model} and \textbf{recommending APIs with the trained model}.
%based on two attentive networks to exploit function intention of method and usage pattern of API on multi-view relational heterogeneous graphs. Figure~\ref{fig:workflow} shows the overall workflow of \tool, which consists of three main stages, including \textbf{constructing multi-view relation graphs}, \textbf{training a graph representation model} and \textbf{recommending APIs with the trained model}. 
In the first stage, we construct three 
% relational
heterogeneous graphs, i.e., \textit{call interaction graph}, \textit{global API co-occurrence graph}, and \textit{hierarchical structure graph}. The nodes of these graphs include APIs, methods, classes, projects, and packages. We extract the relations between nodes from the source code.
% while their corresponding relations are extracted from the source code.
Then, in the second stage, 
% we train 
a graph representation model is proposed to encode the three graphs and integrate the graph representations for recommendation.
% above respectively for recommendation.
In the last stage, we employ the trained model to return a ranked list of API usage recommendation according to the code snippet of the current client method. 
The details of our approach are explained in the following parts.
For facilitating readers' understanding of the proposed approach, we list the key notations
% used 
in Table \ref{tab:symDes}.

% \subsection{Heterogeneous Relation Graphs Construction}
\subsection{Constructing Multi-View Heterogeneous Graphs}\label{subsec:multi_view_graph}

\begin{algorithm}[htb]  
  \caption{{Global API Co-occurrence Graph Construction}($S$, $\varepsilon$)}  
  \label{alg:build_graph}
    \small
    \SetKwInOut{Input}{input}\SetKwInOut{Output}{output}
    \Input {An API sequence set $S$ and an integer $\varepsilon$}
    \Output{A global API co-occurrence graph}
 
    Let $V$ and  $E$ represent the set of vertices and edges in the co-occurrence graph respectively\;
    $V \gets \emptyset$, $E \gets \emptyset$\;
    Let $\omega$ represent a weight function : $E \rightarrow \mathbb{R}^{\geq 0}$, where each edge $e$ in $E$ has a weight $\omega(e)$\;
    % $\mathcal{G}_C \gets $
    \ForEach {$A$ \textbf{in} $S$} {
        % $V \gets V \cup A$\;  
        append all API nodes in $A$ into $V$ \;
        \ForEach{$a_i$ \textbf{in} $A$ }{
          $E \gets E \cup \{(a_i,a_j)|a_j\in{A} \wedge j \in \{j, j+\varepsilon\} \}$ \;
        
         }
    }
    \ForEach{$e$ \textbf{in} $E$ }{
        $\omega(e) \gets \omega(e) + 1$ \;
    }
    $\mathcal{G}_C \gets$ build co-occurrence graph with $V$, $E$ and $\omega$ \;
    \Return $\mathcal{G}_C$\;
    
\end{algorithm}

In this section, we present the graph construction process of the multi-view graphs.
% the process of constructing these three graphs in detail. 
Following the formulation of mainstream recommendation systems~\cite{DBLP:conf/loca/Chen05}, we treat all methods and APIs in projects as \textit{user} set ${\mathcal{U}}$, and \textit{item} set $\mathcal{I}$. The graphs are constructed from local view (i.e., method-API interaction), global view (i.e., API co-occurrence information), and external view (i.e., hierarchical structure), respectively.

\textbf{1) Call Interaction Graph $\mathbf{\mathcal{G}_I}$.} 
It represents the call relations between methods and APIs, denoted as a bipartite graph $\mathcal{G}_I=\{(u,y_{ui},i)| u\in \mathcal{U},i\in\mathcal{I}\}$, where $y_{ui}=1$ indicates a method $u$ calls an API $i$.
%, otherwise $y_{ui}=0$. 
For example, \textit{[myFile.createFile(), 1, java.io.File.exists()]} indicates that a method \textit{myFile.createFile()} calls an API \textit{java.io.File.exists()}. The call interaction graph reflects the basic relations between APIs and methods, and is commonly adopted by prior studies~\cite{DBLP:conf/icse/NguyenN15, DBLP:conf/kbse/Gu0019, DBLP:conf/sigsoft/FowkesS16, DBLP:conf/icse/NguyenRRODP19, DBLP:conf/wcre/LingZX21}.
% is the basic information source. 
% \yujia{By constructing such a graph, we can incorporate historical interaction (i.e., methods who have interacted with the same APIs and APIs that a method has interacted with in history) to obtain the match probability between methods and APIs more accurately.}

%By constructing such a graph, we can learn the historical call information of methods and APIs when recommending.

% \xiaoxue{\textbf{[NEED to explain why we need to construct such graph (maybe we can learn relations between functionality and APIs. For instance, for a client method createFile developers may need to use APIs such as java.io.File.exists() and so on?? use an example pls..) and its role with other graphs...]}}
%In an API recommendation scenario, there exists call relations between method and API. Here we represent these interaction data as a bipartite graph ${\mathcal{G}_I=\{(u,y_{ui},i)| u\in}$
%${\mathcal{U},i\in\mathcal{I}\}}$, where $y_{ui}=1$ indicates method $u$ has called API $i$, otherwise $y_{ui}=0$. 

\textbf{2) Global API Co-occurrence Graph $\mathbf{\mathcal{G}_C}$.} It records the co-occurrence relations between APIs, e.g., the two API \textit{file.open()} and \textit{file.close()} are connected since they ever appeared together in some methods.
% algorithm des.
Algorithm \ref{alg:build_graph} shows the pseudo-code for global API co-occurrence graph construction. The graph is built
% which builds a co-occurrence graph 
based on a set of API sequences $S$ and an integer $\varepsilon$.
Specifically, we first initialize the set of vertices $V$=$\emptyset$ and the set of edges $E$=$\emptyset$ in the co-occurrence graph (line 2).
% \yujia{where $V$ and $E$ represent the set of vertices and edges in the co-occurrence graph, respectively (line 2).}
%\yun{[what are the meaning of the parameters?]}.
%
Then, for each API sequence $A$ in $S$, we append all API nodes in $A$ into $V$ and collect all edges such that each edge $(a_i, a_j)$ s.t. $\forall a_i \in A$ and $\exists j \in \{i, i + \varepsilon\}$, $a_j \in A$ (lines 4-7).
Next, for each edge $e$ in $E$, we update the $\omega(e)$ by counting the occurrence frequencies
% number of times it appears 
(lines 8-9).
Finally, we build the co-occurrence graph $\mathcal{G}_C$ based on $V$, $E$ and $\omega$, and return the co-occurrence graph $\mathcal{G}_C$ (lines 10-11). $\mathcal{G}_C$ is denoted as
% Specifically, the graph is denoted as $\mathcal{G}_C =
$\{(i,f,j)|i,j\in\mathcal{I},f\in\mathcal{T}\}$, where each triplet describes that API $i$ and API $j$ are invocated together $f$ times.
% To construct such graph, we first consider all APIs called by the corresponding method as an API sequence. For $k$-th API ${i_k}$ in a sequence $S$, the $ \varepsilon$-neighbor set of ${i_k}$ is defined as ${\mathcal{N}({i_k}) = \{i_n|i_n\in{S};n\in[k,k+\varepsilon]\}}$, where ${\varepsilon}$ is a window size to control the scope of modeling of API-connection between ${i_k}$ and ${i_n}$. The hyperparameter ${\varepsilon}$ favors the modeling of short-range links since it is helpless (even noise, e.g., irrelevant dependence) for capturing the whole sequence length information if beyond the scope ${\varepsilon}$~\cite{DBLP:conf/sigir/Wang0CLMQ20} \yun{[The long sentence is hard to understand...An algorithm will be better]}. Based on $\varepsilon$-{neighbor} sets of all APIs, we build the API co-occurrence graph between each pair of the APIs, denoted as 
%APIs often appear together several times, we consider frequency ${\mathcal{T}}$ as the edge type of $\mathcal{G}_C$. 
%Formally, the graph is presented as $\{(i,f,j)|i,j\in\mathcal{I},f\in\mathcal{T}\}$, where each triplet describes that $API_i$ and $API_j$ are invocated together $f$ times. 
For example, \textit{[file.open(), 10, file.close()]} indicates that \textit{file.open()} and \textit{file.close()} appear together 10 times. The co-occurrence graph is beneficial for enriching the APIs with frequency information, which can also implicitly complementing the representations of methods.
% By constructing the co-occurrence graph, we can learn the potential co-occurrence information of methods and APIs when recommending.
% [how can it help when recommending APIs?

\textbf{3) Hierarchical Structure Graph $\mathbf{\mathcal{G}_H}$.} The hierarchical information, e.g., the belonging projects/packages, implies the functionality of APIs and methods, thereby helpful for API recommendation.
\begin{figure}
  \centering
  \includegraphics[width=0.45\textwidth]{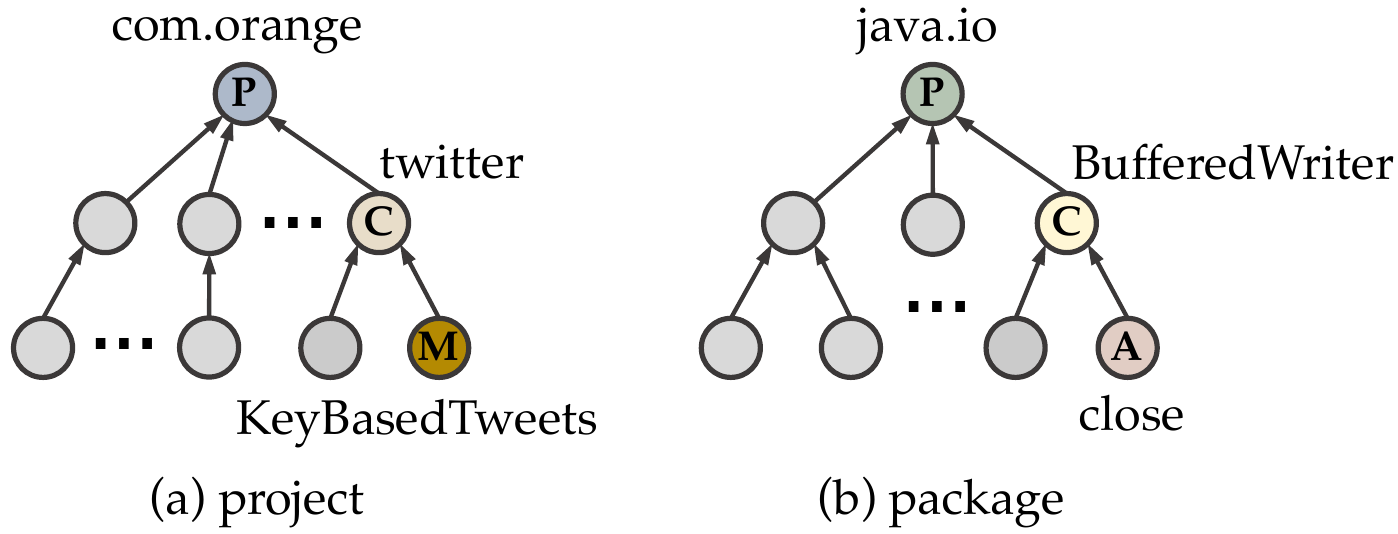}
  \caption{Examples of \textit{Hierarchical Structure Graph}.} %in terms of project and package
  % \xiaoxue{\it{[display which one is project and which is package at the bottom of the %graph]
  \label{fig:graph_exam}
\end{figure}
% Besides call and co-occurrence information, hierarchical information is also essential. 
We consider both project-level and package-level information,
% to construct hierarchical structure graph, 
i.e., the projects where methods are declared and packages that APIs belong to, for constructing the \textit{hierarchical structure graph.}
% \yujia{Project/package level depict the method/API hierarchical attributes, respectively. Methods with similar project structures have more consistent API calls; Likewise, APIs under the same package or class are used together more frequently. Thus, we involve the attribute information by projects and packages, composing the \textit{hierarchical structure graph.}}
%\xiaoxue{\it{[what's the difference between project level and package level? and why we need to consider both two? ]}}
%To construct hierarchical structure graph, We consider both project level and package level information, i.e., the project where a method is declared and the package that API belongs. 
%For example, the tree organizational structure of a project: a project contains plenty of classes, and a class includes some methods. The package structure is similar. By constructing the hierarchical structure graph, we can learn the potential preference information of methods and APIs when recommending.
% Inspired by~\cite{DBLP:journals/tnn/JiPCMY22}, We organize the information in the form of 
We construct the graph as a directed graph, denoted as $\mathcal{G}_H=\{(h,r,t)|h,t\in\mathcal{E},r\in\mathcal{R}\}$, in which each triplet $(h,r,t)$ represents there is a relation $r$ from head entity $h$ to tail entity $t$, $\mathcal{E}$ is the set of all entities, including \textit{API}, \textit{method}, \textit{class}, \textit{project}, and \textit{package}, and $\mathcal{R}$ is the set of relations including \textit{belong-to-class}, \textit{belong-to-project} and \textit{belong-to-package}.
% figure depict
As the example depicted in Figure~\ref{fig:graph_exam}, 
% both 
projects and packages are organized as a tree structure.
% have a tree organizational structure. 
% For example, 
In the example shown in Figure~\ref{fig:graph_exam} (a), \textit{[keyBasedTweents(), belong-to-class, twitter]} denotes that the method \textit{keyBasedTweents()} belongs to the class \textit{twitter}; \textit{[twitter, belong-to-project, com.orange]} denotes that the class \textit{twitter} belongs to the project \textit{com.orange}.
% \yujia{For example, \textit{[write, belong-to-class, PrintWriter]} in the package level denotes that the API \textit{write} belongs to the java class \textit{PrintWriter}; \textit{[processOut, belong-to-class, ParkingManagementController]} in the project level denotes the method \textit{processOut} belongs to the project class \textit{ParkingManagementController}} \yun{[use a figure to illustrate???]}.
%\xiaoxue{\it{[add a project level example here]}} 
% By constructing the hierarchical structure graph, we can learn the potential preference information of methods and APIs when recommending.
% \yun{[add the reason to involve such information?]}
% By constructing the hierarchical structure graph, we can incorporate external contextual information, (i.e., the project and package where the methods and APIs are currently located) to measure semantic match between methods and APIs better when recommending.
% \xiaoxue{\it{[the last sentences of the three graphs ``by constructing XXX graph'' are too general, please make it more specific in recommending]}}
% \subsection{Function Intent and Usage Pattern Mining}

\subsection{Training Graph Representation Model and Recommendation} \label{subsec:encoding}
% After constructing graphs, we need to represent such graph information for serving API recommendation tasks. 
This section introduces how \tool trains a graph representation model based on the constructed multi-view graphs, and utilizes the trained model to make API recommendation, corresponding to the second stage and third stage in Figure~\ref{fig:workflow}, respectively.
% i.e., the second and third stages in Figure~\ref{fig:workflow}.

% \xiaoxue{[This paragraph should be rephrased, I'll revise it in the morning.. We should also state why we need to encode cilent method and target APIs respectively in this paragraph. ]}

Figure~\ref{fig:model_framework} illustrates the whole process of training and recommendation,
% recommending, 
including three graph encoding modules, i.e., \textit{call interaction encoding}, \textit{co-occurrence information encoding} and \textit{hierarchical structure encoding}, as well as one fusion and prediction module. Given a client method, a target API, and the three heterogeneous graphs as input,
% the call interaction graph \yun{[symbol]}, the global API co-occurrence graph and the hierarchical structure graph as input, 
the graph representation model aims to predict the probability of the target API invocated by the client method.
% The whole process mainly contains four modules. 
In the first module, an \textit{Embedding Network} is employed to encode basic interaction information into local-view representations of the client method and the target API, as shown in Figure~\ref{fig:model_framework} (1). Then, in the second module, as illustrated in Figure~\ref{fig:model_framework} (2), a \textit{Frequency-aware Attentive Network} is designed to encode frequency-based co-occurrence information into
% global-view 
representations of the client method and the target API from global view. Next, in the third module, a \textit{Structure-aware Attentive Network} is designed to encode structure-based hierarchical information into 
% external-view 
representations of the client method and the target API from external view, as shown in Figure~\ref{fig:model_framework} (3). Finally, in the last module, the local-view, global-view and external-view representations are concatenated as
% into 
the final representations of the client method and the target API. 

\subsubsection{Call Interaction Encoding} \label{subsection:his encoding}
% 1. module purpose
% 2. module design [follow algorithm]

Call interaction reflects basic information of the client method $u$ and the target API $i$, respectively. We utilize them to generate local-view representations following the prior study
% similar to~
\cite{DBLP:conf/kdd/ZhouZSFZMYJLG18}. Specifically, for each
% the 
client method $u$, the called API set
% that it called 
is denoted as $\mathcal{E}_u = \{i \ | \ i\in\{i \ | \ (u,y_{ui},i) \in \mathcal{G_I} \ and\  y_{ui}=1\}\}$. We then obtain the client method $u$ representation according to its called API set: $ e_u^{(o)} = \frac{ \sum\nolimits_{i \in \mathcal{E}_u} e_i }{|\mathcal{E}_u|}$, where $e_i$ is the embedding of API $i$ and $|\mathcal{E}_u|$ is the set size. 
% In the same way, 
Similarly, we obtain the target API $i$ representation $ e_i^{(o)}$.

\subsubsection{Co-occurrence Information Encoding}\label{subsec:Co-occurrence_Information_Encoding}

\begin{figure}[htp]
\centering
\subfigure[Frequency-aware Attentive Network]{ \label{fig:model_network_f}
\includegraphics[scale=0.4]{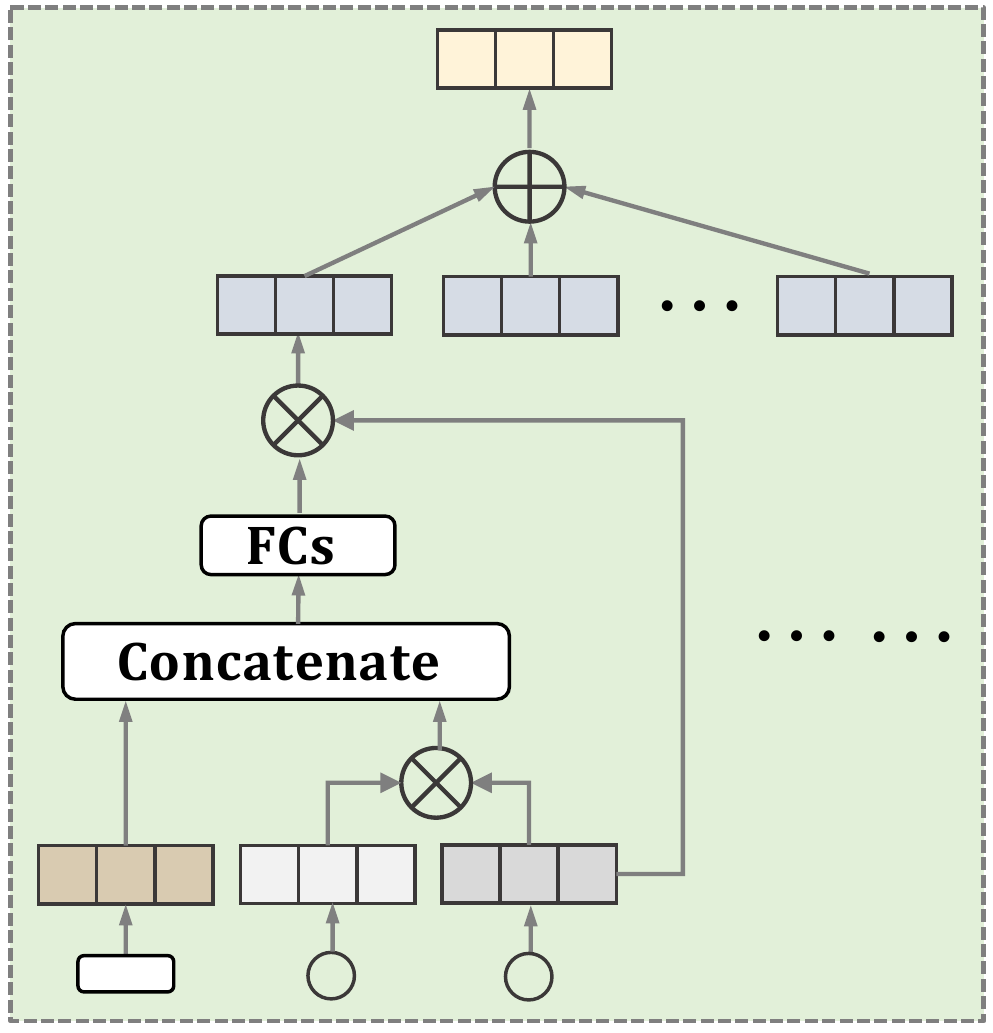}} 
\quad
\centering
\subfigure[Structure-aware Attentive Network]{ \label{fig:model_network_s}
\includegraphics[scale=0.4]{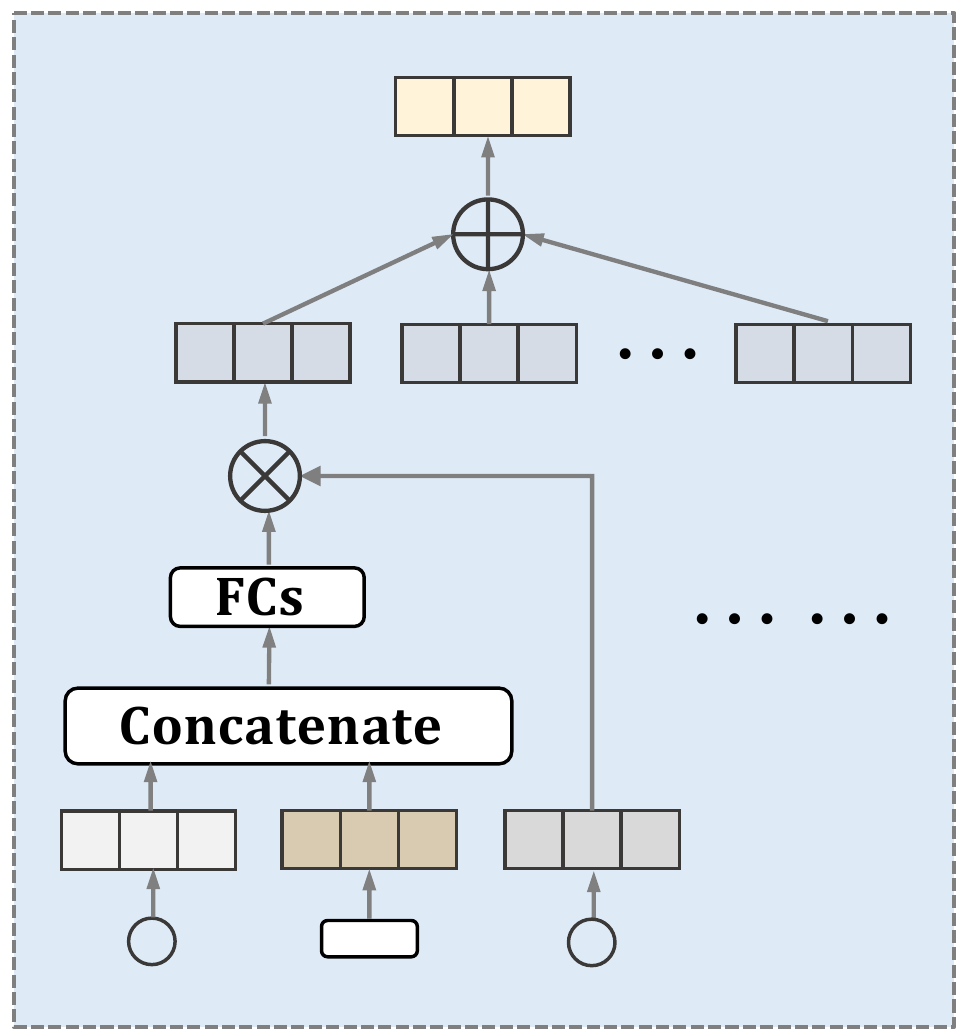}} 
\caption{Illustration of two attentive networks in the graph representation model.}
\label{fig:model_network}
\end{figure}

Global co-occurrence reflects the frequency-enriched information of the client
method $u$ and the target API $i$. According to the definition of \textit{global API co-occurrence graph} in Section~\ref{subsec:multi_view_graph}, the value of the edge between a pair of API nodes denotes their co-occurring frequency, which also implies their relevant connection. 
% Thus, in this module, 
To encode the global-view information of the client method $u$ and the target API $i$, we design a \textit{frequency-aware attentive network}, 
% to encode the information into global view representations of the client method $u$ and the target API $i$, 
as shown in Figure~\ref{fig:model_network_f}.
Algorithm~\ref{alg:encoding_algorithm} shows the pseudo-code
for the encoding. For the client method $u$, the APIs co-occurred with its called APIs reveal the method's potential call need.
% to a certain extent. 
Thus, we utilize the API set $\mathcal{E}_u$ obtained from section~\ref{subsection:his encoding} as the initial set $\mathcal{E}_u^{(0)}$ for the first-hop propagation on $\mathcal{G}_C$ (line 2). After initialization, we conduct
% take 
information encoding to generate co-occurrence representation in each hop.(line 4-8).

\begin{algorithm}[h]  

\caption{Encoding Process of $\mathcal{G}_C$ and $\mathcal{G}_H$}  
  \label{alg:encoding_algorithm}
    \small
    \SetKwInOut{Input}{input}\SetKwInOut{Output}{output}
    \Input {a constructed graph $\mathcal{G}$, a number of max-hop $L$, \textit{AttentiveNetwork}, an entity set $\mathcal{E}$}
    \Output {a list of representation $K_e$}
    
    let $N(v, \mathcal{G})$ be the set of $v$'s neighbors in $\mathcal{G}$\;
    
    let $\mathcal{E}$ be the first-hop entity set $\mathcal{E}^{(0)}$\;
    
    $K_e \gets \emptyset$\;
    
    \For{$l$=$0,1 \cdots, L$}{
        $S^{(l)} \gets \{(i, e(i,j), j) | i \in \mathcal{E}^{(l)} \wedge j \in N(i, \mathcal{G})\}$\;

        $e^{(l)} \gets$  \textit{AttentiveNetwork}($S^{(l)}$)\;
        
        $\mathcal{E}^{(l + 1)} \gets \{\ j \ | i \in \mathcal{E}^{(l)} \wedge j \in N(i, \mathcal{G})\}$\;
    
        % $\mathcal{E}_u^{(l + 1)} \gets$ a set of APIs s.t. each API is a neighbor of the API in $\mathcal{E}_u^{(l)}$ in $\mathcal{G}$ \; 
        append $e^{(l)}$ into $K_e$\;
        
    }
    \Return $K_e$\;

\end{algorithm}

\textbf{Information Encoding (line 4-8).} For each triple in global API co-occurrence graph, i.e., $(i,f,j) \in \mathcal{G}_C$, we define the $l$-hop triple set based on the entity set $\mathcal{E}_u^{(l)}$ as: $\mathcal{S}_u^l = \{(i,f,j)|i \in \mathcal{E}_{u}^{l}\}$ ($l$ begins with 0). Following previous
works~\cite{DBLP:conf/cikm/WangZWZLXG18, DBLP:conf/sigir/WangLTCL20}, we sample a fixed-size triple set instead of using a full-size set to reduce the computation overhead.

Based on sampled co-occurrence associations, i.e., $(i,f,j) \in {S}_u^l$, we learn $l$-hop co-occurrence representation of the client method $u$:
\begin{equation} \label{eq.attention_1}
   e_u^{(l)} = \sum_{(i,f,j) \in \mathcal{S}_u^l} \pi (i,f,j)\ e_j,
\end{equation} 
where coefficient $\pi (i,f,j)$ is attentively calculated as: 
\begin{equation} \label{eq.attention_2}
    \pi (i,f,j) = \frac{\mathrm{exp}(\mathrm{mlp}(((e_i \odot e_j) || e_f))}{\sum_{(i',f',j') \in {\mathcal{S}_u^l}} \mathrm{exp}(\mathrm{mlp}((e_{i'} \odot e_{j'}) || e_{f'}))},
\end{equation}
where the notation $\odot$ denotes the element-wise multiplication operation, and $||$ denotes the concatenation operation. $e_i$ and $e_j$ are the embeddings of API $i$ and its co-occurred API $j$, respectively. $e_f$ is the embedding of frequency $f$. mlp(·) is a three-layer MLP with Relu~\cite{agarap2018deep} as the nonlinear activation function. The attention mechanism for
% in 
encoding the $l$-hop co-occurrence representation (i.e., Equ.~(\ref{eq.attention_1}) and (\ref{eq.attention_2})) explicitly introduces co-occurred frequency $f$ into calculating the influence of co-occurred API $i$ on the representation of API $j$.

% makes the relevance calculation of co-occurred API $j$ depend on the semantic similarity between
% % of 
% API $i$ and API $j$ and co-occurred frequency $f$. 
% For example, co-occurrence triplets (\textit{set.size(), 13, set.add()}) and (\textit{set.isSet(), 5, set.add()}) present two different usage patterns of API \textit{``set.add()''}, which reflects its relevance in representation encoding \yun{[what does this mean??]}.

After performing
% the max hop number, i.e., 
$L$-hop information encoding, where $L$ is the max hop number, we obtain the global-view representations $e_u^{(c)}$ of the client method $u$ by appending the representations from all hops:
% appending all-hop representations: 
$e_u^{(c)} = \{e_u^{(0)}, e_u^{(1)}, \cdots, e_u^{(L)} \}$. Similarly,
% In the same way, 
we
% also 
obtain the global-view representation $e_i^{(c)}$ of the target API $i$.
% [or detailed process?] 
The global-view representations $e_u^{(c)}$ and $e_i^{(c)}$ captures
% collect 
frequency-enriched co-occurrence information for
% , 
enhancing the semantic representations of the client method and the target API, respectively.

\subsubsection{Hierarchical Structure Encoding}
% Hierarchical structure graph usually contains abundant project structure information. For example, from Figure x, we can know method $u_1$'s class is $c_1$ and it is declared in project $p_1$. With the propagation of the initial set on $\mathcal{G}_H$, the structure-based high-order relation information of method and API is successfully captured. For example, method $u_1$ has called $i_1$ and $i_3$, which is considered as initial set. After, $i_1$ propagates to ${c_4}$ and $i_3$ propagates to ${c_3}$ along the links in $\mathcal{G}_H$. From $\mathcal{G}_H$, we can obtain that ${c_3}$ is the class of ${i_2}$. As a result, $u_1$ may have a potential call interaction with ${i_2}$ for $i_2$ belongs to the same class with $i_3$. To characterize methods’ hierarchically extended relation in terms of project structure graph, we recursively define the set of l-hop relevant entities and l-hop entity set for method $u$ similar to $eq.x$ as follows :
% [need rewrite...]

Hierarchical structure reflects external contextual information of the client method $u$ and the target API $i$. According to the definition of \textit{hierarchical structure graph}, as described in Section~\ref{subsec:multi_view_graph}, different head entities and relations endow tail entities with different semantics.
% have diverse meanings for tail entities. 
% Thus, 
To obtain the representations of the client method $u$ and the target API $i$ from external view, we design a \textit{structure-aware attentive network},
% to encode the information into to external view representations of the client method $u$ and the target API $i$, 
as shown in Figure~\ref{fig:model_network_s}.

The encoding process for the hierarchical structural graph is similar to the encoding process of the API co-occurrence information, as illustrated in Section~\ref{subsec:Co-occurrence_Information_Encoding}, except for the design of the attentive network. Specifically,
% shows the pseudo-code for the structure encoding process that is similar to information encoding in section~\ref{subsec:Co-occurrence_Information_Encoding}, except for the attention network. 
for the $l$-hop triple set $\mathcal{S}_u^l = \{(h,r,t)|h \in \mathcal{E}_{u}^{l}\}$ in \textit{hierarchical structure graph}, we learn $l$-hop structure representation for 
% the 
client method $u$ by:
\begin{equation} \label{eq.attention_3}
  e_u^{(l)} = \sum_{(h,r) \in \mathcal{S}_u^l} \pi (h,r)\ e_t,
\end{equation} 
where coefficient $\pi (h,r)$ is attentively calculated as: 
\begin{equation} \label{eq.attention_4}
    \pi (h,r) = \frac{\mathrm{exp}(\mathrm{mlp}(e_h || e_r ))}{\sum_{(h',r',t') \in {\mathcal{S}_u^l}} \mathrm{exp}(\mathrm{mlp}(e_{h'} || e_{r'} ))},
\end{equation}
where $e_h$, $e_t$ are the embeddings of head entity $h$ and tail entity $t$, respectively. $e_r$ is the embedding of relation $r$. The structure-aware attention mechanism (i.e., Equ.~(\ref{eq.attention_3}) and Equ.~(\ref{eq.attention_4})) explicitly endows the relevance calculation of tail entity $t$ with the
% semantic of tail entity $t$ and their
relation $r$.
% makes the relevance calculation of tail entity $t$ depend on the semantic of API $j$ and their relation $r$. 
Based on the structure encoding, we finally
% perform $L$-hop structure encoding to 
obtain the external-view representations $e_u^{(h)}$ and $e_i^{(h)}$ for the client method $u$ and the target API $i$, respectively.

\subsubsection{Fusion and Prediction}

% The representations obtained in each encoding module emphasize different information in source code, 
We obtain
% adopt the $concat$ aggregator~\cite{DBLP:conf/nips/HamiltonYL17} to 
% concatenate the multi-view
% % multiple 
% representations into 
the final representation for the client method $u$ and the target API $i$ by concatenating the multi-view representations, i.e., $e_u = e_u^{(o)} || e_u^{(c)} || e_u^{(h)}$ and $e_i = e_i^{(o)} || e_i^{(c)} || e_i^{(h)}$.
The final representations $e_u$ and $e_i$ incorporate call interaction information, frequency-enriched co-occurrence information and structure-based hierarchical information, 
% which is beneficial 
for accurately capturing
% measuring 
the similarly between the client method $u$ and the target API $i$. 
% Thus, 
During prediction, we conduct inner product of $e_u$ and $e_i$ for calculating
% with its output as 
the call probability: $\hat{y}_{ui}=e_u^\top e_i$.

\section{Experimental Setup}\label{sec:setup}
    In this section, we conduct extensive experiments to evaluate the proposed approach with the aim of answering the following research questions:
\begin{itemize}[leftmargin=*]

\item \textbf{RQ1:} How does \tool perform compared with the state-of-the-art API usage recommendation approaches?
\item \textbf{RQ2:} What is the impact of the three encoding components (i.e., \textit{Call Interaction Encoding, Co-occurrence Information Encoding} and \textit{Hierarchical Structure Encoding}) in the graph representation model on the performance of \tool?
%How do three encoding components of the graph representation model in \tool affect \tool?
\item \textbf{RQ3:} How does \tool perform on low-frequency APIs?
%Does \tool work for APIs with low occurrence frequencies?
\item \textbf{RQ4:} How do different hyper-parameter settings affect \tool's performance?

\end{itemize}

\subsection{Dataset Description}
To evaluate the effectiveness of \tool, we utilize three publicly available benchmark datasets: $SH_S$, $SH_L$, and $MV$:

\begin{itemize}[leftmargin=*]

\item $\mathbf{SH_L}$ contains 610 java projects, filtered from 5,147 randomly downloaded java projects retrieved from GitHub via the Software Heritage archive \cite{DBLP:conf/ipres/CosmoZ17}. 
\item $\mathbf{SH_S}$ is comprised of 200 java projects with small file sizes extracted from $SH_L$. It is designed to evaluate some time-consuming baselines such as PAM~\cite{DBLP:conf/sigsoft/FowkesS16}.
%It is formed aiming at evaluating PAM~\cite{DBLP:conf/sigsoft/FowkesS16} fast, since PAM has a very time-consuming execution process~\cite{DBLP:conf/icse/NguyenRRODP19} and thus is hard to scale on a large dataset.
\item $\mathbf{MV}$ consists of 868 JAR archives collected from the Maven Central repository. There are 3,600 JAR archives in the original dataset, and 1,600 JAR archives remain after being deduplicated by the previous work~\cite{DBLP:conf/icse/NguyenRRODP19, DBLP:conf/wcre/LingZX21}. While through our manual inspection,  we find that the cleaned dataset still contains highly similar projects.
% Note that, the $MV$ is different from the original $MV$ in \cite{DBLP:conf/icse/NguyenRRODP19}. Because through manual inspection, we find that the original $MV$ still contains highly similar projects \yun{after filtering the same projects with different versions out [?].}
% since only the same projects with different version are considered when processing 3,600 JAR archives. Actually, 
For example, some projects have snapshot versions during the development process and a release version at the end, such as \emph{commons-1.0.2.RELEASE.jar} and \emph{commons-1.0.2.BUILD-SNAPSHOT.jar}. Besides, some projects may have their renamed versions, such as \emph{eclipse.equinox.common-3.6.200.jar} and \emph{common-3.6.200.jar}. In these cases, the two projects are nearly identical. Too many similar projects in a dataset may introduce bias in evaluation \cite{DBLP:conf/icse/NguyenRRODP19}. Therefore, we decided further clean
% rebuild 
this dataset by removing duplicated project versions, i.e., the projects with snapshot versions or renamed versions. We finally obtain 868 JAR archives from 3,600 JAR archives for the $MV$ dataset.
% To be specific, we select one version and remove the rest for projects that meet the three conditions above, and obtain 868 JAR archives from 3600 JAR archives as $MV$ finally.
\end{itemize}

\begin{table}
  \centering
  \normalsize
  \caption{Statistics of the three datasets: $\mathbf{SH_S}$, $\mathbf{SH_L}$ and $\mathbf{MV}$. The Call-Avg means the average calls per method.}
  \label{tab:dataDes}
  \begin{tabular}{c|ccc}
    \Xhline{1pt}
       & $SH_S$ & $SH_L$ & $MV$ \\
    \Xhline{0.5pt}
    \# Projects  & 200 & 610 & 868\\
    \# Packages  & 253 & 714 & 340\\
    \# Classes   & 4,285 & 91,060 & 23,207\\  
    \# Methods  & 4,530 & 191,532 & 32,987\\
    \# APIs & 5,351 & 30,576 & 22,054\\
    \# Calls  & 27,312 & 1,027,644 & 343,010\\
    \# Call-Avg & 6 & 5 & 10\\
    \Xhline{1pt}
 \end{tabular}
\end{table}

From the source code in datasets, we extract the method declarations and corresponding API calls, and hierarchical structure of the projects and packages containing methods/APIs.
% which methods are declared in and APIs belong to, respectively. 
We summarize the detailed statistics of the three datasets in Table~\ref{tab:dataDes}.

\subsection{Baselines}
To demonstrate the effectiveness, we compare our proposed \tool with one statistic-based method (PAM), two collaborative filtering(CF)-based methods (FOCUS and GAPI), as follows:

\begin{itemize}[leftmargin=*]

\item \textbf{PAM}~\cite{DBLP:conf/sigsoft/FowkesS16} is a statistical method to mine API usage patterns, which mainly adopts a probabilistic model to infer the patterns with the highest probabilities from client code. 
\item \textbf{FOCUS}~\cite{DBLP:conf/icse/NguyenRRODP19} leverages collaborative filtering to recommend API usage patterns. It measures the similarity between methods via a context-based rating matrix to recommend potential APIs.
%is based on the classical collaborative filtering idea, which measures the similarity between methods by a context-based rating matrix to recommend potential APIs. 
\item \textbf{GAPI}~\cite{DBLP:conf/wcre/LingZX21} is the state-of-the-art CF-based method that employs graph neural networks to capture high-order connectivity between methods and APIs from a unified graph. We re-implement this model according to the original paper. 
\end{itemize}

\subsection{Evaluation Metrics}
Following previous approaches~\cite{DBLP:conf/icse/NguyenRRODP19,DBLP:conf/wcre/LingZX21} on API usage recommendation, we adopt $successRate@K$, $Precision@K$ and $Recall@K$ to evaluate the quality of top-K API usage recommendation.
%and set $K$ to $1$, $5$, $10$, and $20$. 
Given a top-K ranked recommendation list $REC_k(m)$ for a test method $m \in \mathcal{M}$ and the ground-truth set $GT(m)$, we adopt $MATCH_k(m) = REC_k(m) \cap GT(m)$ to present the correctly predicted API set. The $SuccessRate@K(SR@K)$, $Precision@K(P@K)$, and $Recall@K(R@K)$ are defined as follows:
\begin{itemize}[leftmargin=*]

\item $SuccessRate@K$ is the proportion of at least one successful match among the
top-K APIs.
\begin{equation}
    SR@K = \frac{count_{m \in \mathcal{M}}(MATCH_k(m) > 0)}{|\mathcal{M}|}
\end{equation}

\item $Precision@K$ is the proportion of correctly predicted APIs amongst the top-K APIs.
\begin{equation}
    P@K = \frac{|MATCH_k(m)|}{k}
\end{equation}

\item $Recall@K$ is the proportion of correctly predicted APIs amongst the ground-truth APIs. 
\begin{equation}
    R@K = \frac{|MATCH_k(m)|}{|GT(m)|}
\end{equation}

\end{itemize}

\subsection{Implementation Details}

Following the real development scenario settings simulated by FOCUS, we take the last method of each project as the test method, i.e., the active method that the developer is working on. The first four APIs in this method are considered as visible context, added to the \emph{training set,} and the rest is used as ground truth, added to \emph{testing set}. 

We implement \tool in PyTorch. The embedding size is set to 64 for the model in \tool. We employ the binary cross-entropy loss as the loss function. To initialize the model parameters, we utilize the default Xavier initializer~\cite{DBLP:journals/jmlr/GlorotB10}. Also, we choose Adam optimizer~\cite{DBLP:journals/corr/KingmaB14} to train our model, with a learning rate of 0.002, a coefficient of $L2$ normalization of $10^{-5}$, a batch size of 1024 and an epoch number equal to 40 fixed for all datasets. 

Following previous work~\cite{DBLP:conf/sigir/Wang0CLMQ20}, we set the maximum distance of adjacent APIs $\epsilon$ as 3 in constructing $\mathcal{G}_C$. Considering that the edge attribute in $\mathcal{G}_C$ is a continuous variable, we
% Besides, the attribute of the edge in the $G_C$ is frequency, a continuous variable, so we 
adopt the equidistant bucket discretization method, and regard the bucket number as the edge type.
% the number of buckets represents the number of edge types. 
The optimal number of buckets $|\mathcal{T}|$ in discretization, the max hop number $L$ and the size of triple set $|\mathcal{S}_u^l|$ in each hop $l$ on three datasets are determined based on the experimental performance. %\yun{[to check?]}. 
The best settings of the hyper-parameters for all the baseline approaches are defined following the original papers. All approaches are trained on NVIDIA Tesla V100 GPU.

\section{Results}\label{sec:results}

\begin{table*}[htbp] 
  \caption{The performance comparison of $SR@K$ between \tool and the baselines on three datasets.}
  \label{tab:res-topk_SR}
  \begin{center}
  \begin{tabular}{c|cccc|cccc|cccc}
    \toprule
    \multirow{2}{*}{Method} & \multicolumn{4}{c}{$SH_S$} & \multicolumn{4}{|c}{$SH_L$} & \multicolumn{4}{|c}{$MV$} \\ 
    & \textit{SR@1} & \textit{SR@5}    & \textit{SR@10} & \textit{SR@20}  & \textit{SR@1} & \textit{SR@5} & \textit{SR@10} & \textit{SR@20}  & \textit{SR@1} & \textit{SR@5} & \textit{SR@10} & \textit{SR@20} \\
    \midrule
    \midrule
PAM    & 0.080 & 0.150 & 0.275 & 0.335 &-       &-        &-        &-       &-      &-       &-       &-       \\
FOCUS  & 0.161 & 0.256 & 0.328 & 0.422 & 0.188 & 0.292 & 0.349 & 0.388 & 0.549 & 0.709 & 0.769 & 0.819 \\
GAPI & 0.195 & 0.363 & 0.479 & 0.600 & 0.163 & 0.402 & 0.532 & 0.670 & 0.260 & 0.569 & 0.714 & 0.837 \\
    \midrule
    \midrule
    \tool    & \textbf{0.439} & \textbf{0.672} & \textbf{0.794} & \textbf{0.836} & \textbf{0.334} & \textbf{0.544} & \textbf{0.641} & \textbf{0.731} & \textbf{0.658} & \textbf{0.810} & \textbf{0.840} & \textbf{0.875} \\
  \bottomrule
 \end{tabular}
\end{center}
\end{table*}

\begin{table*}[htbp]
  \caption{The performance comparison of $SR@K$ between \tool and its variants on three datasets.}
  \label{tab:res-aba}
  \begin{center}
  \begin{tabular}{c|cccc|cccc|cccc}
    \toprule
    \multirow{2}{*}{Method} & \multicolumn{4}{c}{$SH_S$} & \multicolumn{4}{|c}{$SH_L$} & \multicolumn{4}{|c}{$MV$} \\ 
    & \textit{SR@1} & \textit{SR@5}    & \textit{SR@10} & \textit{SR@20}  & \textit{SR@1} & \textit{SR@5} & \textit{SR@10} & \textit{SR@20}  & \textit{SR@1} & \textit{SR@5} & \textit{SR@10} & \textit{SR@20} \\
    \midrule
    \midrule
\tool w/o H\&C & 0.333     & 0.566     & 0.688     & 0.751     & 0.142     & 0.291     & 0.413     & 0.558     & 0.521     & 0.694     & 0.756     & 0.802\\
    \tool w/o HS & 0.349 & 0.614 & 0.704 & 0.773 & 0.167 & 0.316 & 0.429 & 0.578 & 0.533 & 0.704 & 0.784 & 0.842  \\
    \tool w/o CO & 0.423     & 0.651     & 0.757     & 0.815    & 0.311    & 0.521    & 0.629    & 0.728    & 0.614     & 0.746     & 0.780     & 0.816     \\
    \midrule
    \midrule
    \tool    & \textbf{0.439} & \textbf{0.672} & \textbf{0.794} & \textbf{0.836} & \textbf{0.334} & \textbf{0.544} & \textbf{0.641} & \textbf{0.731} & \textbf{0.658} & \textbf{0.810} & \textbf{0.840} & \textbf{0.875} \\

  \bottomrule
  
\end{tabular}
\end{center}
\end{table*}

\pgfplotstableread[row sep=\\,col sep=&]{
	k & MERIS & GAPI & FOCUS & PAM  \\
	1 & 0.439 & 0.195 & 0.161 & 0.080  \\
	5 & 0.234 & 0.102 & 0.098 & 0.030 \\
	10 & 0.168 & 0.086 & 0.089 & 0.033 \\
	15 & 0.115 & 0.058 & 0.060 & 0.022 \\
}\SHSPRE

\pgfplotstableread[row sep=\\,col sep=&]{
	k & MERIS & GAPI & FOCUS  \\
	1 & 0.334 & 0.163 & 0.188   \\
	5 & 0.165 & 0.115 & 0.114 \\
	10 & 0.117 & 0.090 & 0.077 \\
	15 &0.077 & 0.068 & 0.046  \\
}\SHLPRE

\pgfplotstableread[row sep=\\,col sep=&]{
	k & MERIS & GAPI & FOCUS  \\
	1 & 0.658 & 0.195 & 0.549  \\
	5 & 0.345 & 0.102 & 0.305\\
	10 & 0.222 & 0.086 & 0.193 \\
	15 & 0.135 & 0.058 & 0.120\\
}\MVSPRE

\pgfplotstableread[row sep=\\,col sep=&]{
	k & MERIS & GAPI & FOCUS & PAM  \\
	1 & 0.102 & 0.044 & 0.039 & 0.013  \\
	5 & 0.244 & 0.112 & 0.098 & 0.028 \\
	10 & 0.331 & 0.169 & 0.167 & 0.057 \\
	15 & 0.420 & 0.232 & 0.214 & 0.071 \\
}\SHSREC

\pgfplotstableread[row sep=\\,col sep=&]{
	k & MERIS & GAPI & FOCUS  \\
	1 & 0.093 & 0.048 & 0.076   \\
	5 & 0.208 & 0.152 & 0.176 \\
	10 & 0.286 & 0.219 & 0.211 \\
	15 & 0.367 & 0.308 & 0.244  \\
}\SHLREC

\pgfplotstableread[row sep=\\,col sep=&]{
	k & MERIS & GAPI & FOCUS  \\
	1 & 0.130 & 0.055 & 0.107  \\
	5 & 0.313 & 0.199 & 0.259\\
	10 & 0.381 & 0.289 & 0.318 \\
	15 & 0.447 & 0.394 & 0.382\\
}\MVSREC

\definecolor{c1}{RGB}{178,37,42} % MERIS
\definecolor{c2}{RGB}{255,196,115} % GAPI
\definecolor{c3}{RGB}{103,150,118} % FOCUS
\definecolor{c4}{RGB}{181,181,181} % PAM

\begin{figure*}[h]
	\centering		
    % \ref{named1} \ref{named2} \ref{named3}\\
	\subfigure[$SH_S$]{
		\begin{tikzpicture}[scale=0.63]
			\begin{axis}[
			    legend style = {
				    legend columns=2,
				    draw=none,
				},
				width=0.46\textwidth,
    			height=0.36\textwidth,
				% legend to name,
				xtick = {1,5,10,15},
				xticklabels = {1, 5, 10, 20},
				ymin=0,ymax=0.5,
				ytick = {0.1, 0.2, 0.3, 0.4},
				mark size=2.5pt, 
				ylabel={\Large \bf P@K},
				xlabel={\Large $\mathbf{K}$}, 
				% ticklabel style={font=\Huge},
				ticklabel style={font=\Large},
				every axis plot/.append style={line width = 2pt},
				every axis/.append style={line width = 1.3pt},
				]
				\addplot [mark=star,color=c1] table[x=k,y=MERIS]{\SHSPRE};
				\addplot [mark=triangle*,color=c2] table[x=k,y=GAPI]{\SHSPRE};
				\addplot [mark=pentagon*,color=c3] table[x=k,y=FOCUS]{\SHSPRE};
				\addplot [mark=diamond*,color=c4] table[x=k,y=PAM]{\SHSPRE};
				
				% \addlegendentry{$x$}
				\legend{MEGA, GAPI, FOCUS, PAM}
			\end{axis}
		\end{tikzpicture}
		\label{fig:res/pre.SHS}
	}
	\subfigure[$SH_L$]{
		\begin{tikzpicture}[scale=0.63]
			\begin{axis}[
			   legend style = {
				    legend columns=2,
				    draw=none,
				},
				ymin=0,ymax=0.45,
				ytick = {0.1, 0.2, 0.3, 0.4},
				xtick = {1,5,10,15},
				xticklabels = {1, 5, 10, 20},
				mark size=2.5pt, 
				width=0.46\textwidth,
    			height=0.36\textwidth,
				ylabel={\Large \bf P@K},
				xlabel={\Large $\mathbf{K}$}, 
				ticklabel style={font=\Large},
				every axis plot/.append style={line width = 2pt},
				every axis/.append style={line width = 1.3pt},
				]
				\addplot [mark=star,color=c1] table[x=k,y=MERIS]{\SHLPRE};
				\addplot [mark=triangle*,color=c2] table[x=k,y=GAPI]{\SHLPRE};
				\addplot [mark=pentagon*,color=c3] table[x=k,y=FOCUS]{\SHLPRE};
				\legend{MEGA, GAPI, FOCUS}
			\end{axis}
		\end{tikzpicture}
    	\label{fig:res/pre.SHL}
	}
	\subfigure[$MV_S$]{
		\begin{tikzpicture}[scale=0.63]
			\begin{axis}[
				ytick = {0.1, 0.2, 0.3,0.4, 0.5, 0.6},
				% width=0.53\textwidth,
    % 			height=.36\textwidth,
    			 legend style = {
				    legend columns=2,
				    draw=none,
				},
				xtick = {1,5,10,15},
				xticklabels = {1, 5, 10, 20},
				ymin=0,ymax=0.7,
				mark size=2.5pt, 
				width=0.46\textwidth,
    			height=0.36\textwidth,
				ylabel={\Large \bf P@K},
				xlabel={\Large $\mathbf{K}$}, 
				ticklabel style={font=\Large},
				every axis plot/.append style={line width =2pt},
				every axis/.append style={line width = 1.3pt},
				]
				\addplot [mark=star,color=c1] table[x=k,y=MERIS]{\MVSPRE};
				\addplot [mark=triangle*,color=c2] table[x=k,y=GAPI]{\MVSPRE};
				\addplot [mark=pentagon*,color=c3] table[x=k,y=FOCUS]{\MVSPRE};
			    \legend{MEGA, GAPI, FOCUS}
			\end{axis}
		\end{tikzpicture}
		\label{fig:res/pre.MV}
	}
	\caption{The performance comparison of $P@K$ between \tool and the baselines on three datasets.}
	\label{fig:res-topk_pre}
% 	\vspace{-0.3cm} 
\end{figure*}
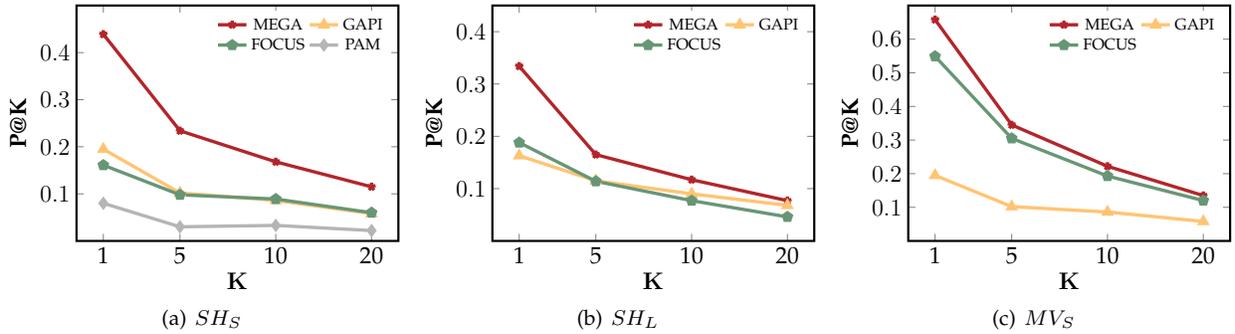

\begin{figure*}[h]
	\centering		
    % \ref{named1} \ref{named2} \ref{named3}\\
    
	\subfigure[$SH_S$]{
		\begin{tikzpicture}[scale=0.63]
			\begin{axis}[
			    legend style = {
				    legend columns=2,
				    draw=none,
				},
				legend pos=north west,   
				width=0.46\textwidth,
    			height=0.36\textwidth,
				% legend to name,
				xtick = {1,5,10,15},
				xticklabels = {1, 5, 10, 20},
				ymin=0,ymax=0.5,
				ytick = {0.1, 0.2, 0.3, 0.4},
				mark size=2.5pt, 
				ylabel={\Large \bf R@K},
				xlabel={\Large $\mathbf{K}$}, 
				% ticklabel style={font=\Huge},
				ticklabel style={font=\Large},
				every axis plot/.append style={line width = 2pt},
				every axis/.append style={line width = 1.3pt},
				]
				\addplot [mark=star,color=c1] table[x=k,y=MERIS]{\SHSREC};
				\addplot [mark=triangle*,color=c2] table[x=k,y=GAPI]{\SHSREC};
				\addplot [mark=pentagon*,color=c3] table[x=k,y=FOCUS]{\SHSREC};
				\addplot [mark=diamond*,color=c4] table[x=k,y=PAM]{\SHSREC};
				
				\legend{MEGA, GAPI, FOCUS, PAM}
			\end{axis}
		\end{tikzpicture}
		\label{fig:res/rec.SHS}
	}
	\subfigure[$SH_L$]{
		\begin{tikzpicture}[scale=0.63]
			\begin{axis}[
			   legend style = {
				    legend columns=2,
				    draw=none,
				},
				legend pos=north west,  
				ymin=0,ymax=0.45,
				ytick = {0.1, 0.2, 0.3, 0.4},
				xtick = {1,5,10,15},
				xticklabels = {1, 5, 10, 20},
				mark size=2.5pt, 
				width=0.46\textwidth,
    			height=0.36\textwidth,
				ylabel={\Large \bf R@K},
				xlabel={\Large $\mathbf{K}$}, 
				ticklabel style={font=\Large},
				every axis plot/.append style={line width = 2pt},
				every axis/.append style={line width = 1.3pt},
				]
				\addplot [mark=star,color=c1] table[x=k,y=MERIS]{\SHLREC};
				\addplot [mark=triangle*,color=c2] table[x=k,y=GAPI]{\SHLREC};
				\addplot [mark=pentagon*,color=c3] table[x=k,y=FOCUS]{\SHLREC};
				\legend{MEGA, GAPI, FOCUS}
			\end{axis}
		\end{tikzpicture}
		\label{fig:res/rec.SHL}
	}
	\subfigure[$MV_S$]{
		\begin{tikzpicture}[scale=0.63]
			\begin{axis}[
				ytick = {0.1, 0.2, 0.3,0.4, 0.5},
				% width=0.53\textwidth,
    % 			height=.36\textwidth,
    			 legend style = {
				    legend columns=2,
				    draw=none,
				},
				legend pos=north west,  
				xtick = {1,5,10,15},
				xticklabels = {1, 5, 10, 20},
				ymin=0,ymax=0.6,
				mark size=2.5pt, 
				width=0.46\textwidth,
    			height=0.36\textwidth,
				ylabel={\Large \bf R@K},
				xlabel={\Large $\mathbf{K}$}, 
				ticklabel style={font=\Large},
				every axis plot/.append style={line width =2pt},
				every axis/.append style={line width = 1.3pt},
				]
				\addplot [mark=star,color=c1] table[x=k,y=MERIS]{\MVSREC};
				\addplot [mark=triangle*,color=c2] table[x=k,y=GAPI]{\MVSREC};
				\addplot [mark=pentagon*,color=c3] table[x=k,y=FOCUS]{\MVSREC};
			    \legend{MEGA, GAPI, FOCUS}
			\end{axis}
		\end{tikzpicture}
		\label{fig:res/rec.MV}
	}
	\caption{The performance comparison of $R@K$ between \tool and the baselines on three datasets.}
	\label{fig:res-topk_rec}
% 	\vspace{-0.3cm} 
\end{figure*}
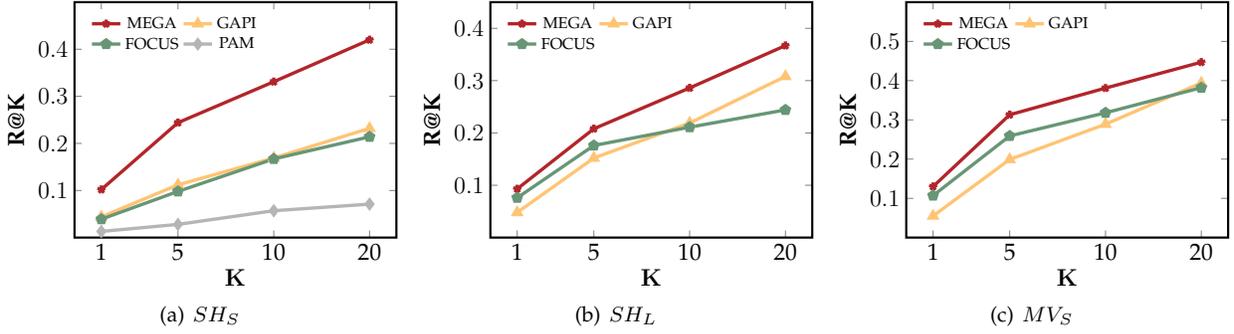

% \begin{figure*}[htp] 
% \centering
% \subfigure[$SH_S$]{\label {fig:res/pre.SHS}
% \includegraphics[scale=0.27]{figures/pre_SH_S.pdf}} 
% \quad
% \subfigure[$SH_L$]{\label {fig:res/pre.SHL}
% \includegraphics[scale=0.27]{figures/pre_SH_L.pdf}} 
% \quad
% \subfigure[$MV$]{\label {fig:res/pre.MV}
% \includegraphics[scale=0.27]{figures/pre_MV_S.pdf}} 
% \caption{The performance comparison of $Precision@K$ between \tool and the baselines on three datasets.}
% \label{fig:res-topk_pre}
% \end{figure*}

% \begin{figure*}[htp]
% \centering
% \subfigure[$SH_S$]{ \label {fig:res/rec.SHS}
% \includegraphics[scale=0.27]{figures/recall_SH_S.pdf}} 
% \quad
% \subfigure[$SH_L$]{ \label {fig:res/rec.SHL}
% \includegraphics[scale=0.27]{figures/recall_SH_L.pdf}} 
% \quad
% \subfigure[$MV$]{ \label {fig:res/rec.MV}
% \includegraphics[scale=0.27]{figures/recall_MV_S.pdf}} 
% \caption{The performance comparison of $Recall@K$ between \tool and the baselines on three datasets.}
% \label{fig:res-topk_rec}
% \end{figure*}

\subsection{Effectiveness of \tool Compared with Baselines (RQ1)}

Table~\ref{tab:res-topk_SR} presents overall results of all baselines along with \tool in terms of $SR@K$ metric, and the comparison curves of $P@K$ and $R@K$ on three datasets (with $K$ = {1, 5, 10, 20}) are shown in Figure~\ref{fig:res-topk_pre} and Figure~\ref{fig:res-topk_rec}, respectively. Intuitively, \tool consistently achieves the best performance on all datasets. Note that, we only test PAM on $SH_S$ due to its long execution time and scaling poorly in a large dataset. Detailed observations are as follows:

\textbf{Comparison of $SR@K$ on a single dataset.} 
%We take the $SH_S$ dataset as an example to illustrate the comparison while a similar trend is also seen in other datasets. 
Without loss of generality, we take the $SH_S$ dataset as an example to illustrate the comparison here, and similar trends can also be observed on other datasets. In the $SH_S$ dataset, \tool improves over the state-of-the-art baseline GAPI $w.r.t$ SR@1, SR@5, SR@10, and SR@20 by 125.1\%, 85.12\%, 65.76\% and 39.33\% respectively. This demonstrates the effectiveness of \tool on various Top-K settings. 
%In addition, shown from the results, SR@K improves as K becomes large. 
What's more, from Table~\ref{tab:res-topk_SR} we can see that $SR@K$ of \tool increases to 0.836 when K increases to 20. This means that in most cases, \tool can identify the correct API in the Top-20 results, while other baseline models can only identify about 60\% of correct APIs in the Top-20 results. 
%Specifically, $SR@K$ of \tool has increased from 0.439 to 0.834, which is almost doubled when K changes from 1 to 10. This is an intuitive result, meaning that the more candidate APIs are provided, the more likely one matched API exists. 

%proving our method applies to various TOP-K recommendation tasks, and achieves outstanding performance.
%\textcolor{red}{[Highlight the percentage of improvement?]}

\textbf{Comparison of $SR@1$ on multiple datasets.} To evaluate the performance of \tool among multiple datasets compared with baseline models, we choose the $SR@1$ metric as it considers both whether a correct API can be included and whether a correct API can get a higher rank. 
%Empirically, if there is only one API in the recommendation list, this API is exactly what is required for the client method, which is more indicative of the proposed approach's strong ability to make accurate recommendations. Therefore, we choose $SR@1$ as the comparison metric.
Overall, in terms of $SR@1$, \tool improves 125.1\% and 77.65\% compared to the best baseline GAPI on $SH_S$ and $SH_L$, and 19.85\% compared to the best baseline FOCUS on $MV$. This demonstrates that \tool is more effective on multiple datasets than baselines. We notice that GAPI underperforms FOCUS on $MV$. One possible explanation is that when the first-order call interactions are abundant enough, high-order connectivity introduces more noise into the representation of methods and APIs instead, leading to a negative effect.

It is worth noting that \tool has more prominent results in $SH_S$, which is the smallest dataset with the minimum number of average interactions for per method. Specifically, the $SR@1$ of FOCUS and GAPI is 0.161 and 0.195, and for PAM, it is even lower at 0.08, which means that all baselines fail to provide the correct API at the first position for more than 80\% of cases.
%which implies that all baselines fail to provide at least one correct API usage for more than 80\% of the client methods. 
Whereas, the same metric for \tool is 0.439, meaning that \tool can successfully recommend the correct API in the Top-1 result for nearly half of the client methods. The significant improvement of \tool on $SR@1$ verifies our approach of encoding diverse information into final representation, especially when historical call interactions are sparse in small dataset.

Different from $SR@K$ and $R@K$ that increases when $K$ increases, we find that $P@K$ decreases when $K$ increases. Because in most cases, the number of correct APIs is much fewer than the candidate list size $K$. While increasing the candidate list size helps \tool find the correct API, it also involves many irrelevant APIs.

%This verifies the significance of encoding diverse information into final representation, especially when historical call interactions are sparse in small dataset. 

%\textcolor{red}{[Highlight the percentage of improvement?]}

%We notice that, the curve of Figure~\ref{fig:res-topk_pre} shows a downward trend, while the curve of Figure~\ref{fig:res-topk_rec} shows an upward trend. One explanation is as the number of candidates ($i.e.$ K) increases, the recall rate also increases, because it is more likely to find the API in ground-truth, which is at the cost of decreased the precision rate. 

%\textbf{Overall Comparison.} 
The above experimental results show that pattern-based methods, such as PAM, relying on mining frequent subsequences generally perform worse than learning-based methods such as GAPI and \tool. This indicates the significance of exploring high-order connections and making full use of external information.

%the traditional methods based on statistics performs worse than the latest methods by a large margin since just relying on mining frequent subsequences is hard to accurately gain API usage patterns, which causes poor performance in providing API usage recommendation. Besides, in most of the scenarios, learning-based methods achieve better performance than pattern-based methods, indicating the significance of exploring high-order connections and making full use of external information.

%To summarize, we conclude that \tool is an effective method that is far superior to the state-of-the-art methods in terms of success rate and accuracy for API usage recommendation. 

% \begin{mybox}
%   To summarize, we conclude that \tool is an effective method that is far superior to the state-of-the-art methods in terms of success rate and accuracy for API usage recommendation.
% \end{mybox}

\begin{tcolorbox}[breakable,width=\linewidth-2pt,boxrule=0pt,top=4pt, bottom=4pt, left=7pt,right=7pt, colback=gray!20,colframe=gray!20]
\textbf{Answer to RQ1:} 
\tool is quite effective on API usage recommendation, outperforming the state-of-the-art approaches in terms of Success Rate, Precision and Recall, respectively. Besides, \tool achieves consistent performance on all datasets. 
%\tool is an effective approach that is far superior to the state-of-the-art approaches in terms of success rate and accuracy for API usage recommendation.
\end{tcolorbox}

\subsection{Ablation Study (RQ2)}
To investigate the effectiveness of each component of the graph representation model in \tool, we perform ablation studies by considering the following three variants.
\begin{itemize}[leftmargin=*]

\item \textbf{$\tool_{\mathrm{w/o\ HS}}$:} This variant removes the hierarchical structure encoding module from the model to study the effect of external information derived from the project and package.
\item \textbf{$\tool_{\mathrm{w/o\ CO}}$:} This variant deletes the co-occurrence information encoding module from the model to investigate the impact of global information obtained between APIs.
\item \textbf{$\tool_{\mathrm{w/o\ H\&C}}$:} This variant only preserves the call interaction encoding in the model to gain the primary representations of the method and the API, without any supplementary information.

\end{itemize}

The experimental results are shown in Table~\ref{tab:res-aba}. We find that the performance of \tool drops in three variants compared with the complete model, which demonstrates the effectiveness of the hierarchical structure encoding module and co-occurrence information encoding module.

$\tool_{\mathrm{w/o\ H\&C}}$ performs worst since the variant only utilizes historical call information. Moreover, we notice that the performance degradation is most significant on $SH_L$. For example, $SR@1$ decreases from 0.311 to 0.142. Note that, in $SH_L$, the average number of call interactions is 5, which is the smallest among all datasets. This demonstrates the prominent advantage of appending co-occurrence information and structure information to the final representation when the interactions are insufficient. 

In addition, the performance of $\tool_{\mathrm{w/o\ HS}}$ is better than $\tool_{\mathrm{w/o\ CO}}$, meaning that hierarchical structure information is more critical than co-occurrence information. One possible reason is that the external project/package structure can provide more contextual information instead of just internal call relation, which is more beneficial for capturing the semantic match between methods and APIs.
% \xiaoxue{[Can we give more explanation with specific cases?]}

\begin{tcolorbox}[breakable,width=\linewidth-2pt,boxrule=0pt,top=4pt, bottom=4pt, left=7pt,right=7pt, colback=gray!20,colframe=gray!20]
\textbf{Answer to RQ2:} Encoding both API co-occurrence information and project/package hierarchical structure information into the final representations of methods and APIs is beneficial to \tool's performance improvements. Especially, project/package hierarchical structure information is more critical, benefiting from it contains contextual information of methods and APIs.
\end{tcolorbox}

\begin{table}
 \small
  \caption{The performance comparison over low-frequency APIs on three datasets.}
  \label{tab:res-few}
  \begin{tabular}{c|cc|cc|cc}
  \toprule
    % \multirow{2}{*}{Model} & \multicolumn{2}{c}{$S$} & \multicolumn{2}{|c}{$SH_L$} & \multicolumn{2}{|c}{$MV$} \\ 
     & \textit{SR@1}    & \textit{SR@10}  & \textit{P@1}  & \textit{P@10}  & \textit{R@1}  & \textit{R@10}  \\
    \midrule
    % \midrule
    \multicolumn{7}{c}{$SH_S$ } \\
    \midrule
FOCUS  & 0.017  & 0.156  & 0.017 & 0.016 & 0.009  & 0.043  \\
GAPI & 0.020  & 0.081  & 0.020 & 0.013 & 0.007  & 0.034  \\
\tool    & \textbf{0.081}  & \textbf{0.263}  & \textbf{0.081}  &  \textbf{0.046}  & \textbf{0.029} & \textbf{0.176}  \\
\midrule
    \multicolumn{7}{c}{$SH_L$ } \\
    \midrule
    FOCUS  & 0.003  & 0.004  & 0.003 & 0.004 & 0.002  & 0.015  \\
GAPI & 0.040  & 0.079  & 0.040 & 0.012 & 0.020  & 0.047  \\
\tool    & \textbf{0.081}  & \textbf{0.315}  & \textbf{0.081}  & \textbf{0.040}  & \textbf{0.057} & \textbf{0.236}  \\
\midrule
    \multicolumn{7}{c}{$MV$ } \\
    \midrule
    FOCUS  & 0.002  & 0.003  & 0.003 & 0.002 & 0.001  & 0.007  \\
GAPI & 0.009  & 0.018  & 0.009 & 0.003 & 0.006  & 0.016  \\
\tool    & \textbf{0.014}  & \textbf{0.041}  & \textbf{0.014}  & \textbf{0.008}  & \textbf{0.006} & \textbf{0.029}  \\
   \bottomrule
\end{tabular}
\end{table}

\subsection{Performance of \tool on low-frequency APIs(RQ3)}

As stated in Section~\ref{sec:motivation}, we design \tool to alleviate the problem that current approaches on low-frequency APIs. To verify our design, we conduct experiments on APIs called by methods less than or equal to 3 times.
% 	\caption{The performance comparison of $Recall@K$ between \tool and the baselines on three datasets.}
% 	\label{fig:res-para-hb}
% % 	\vspace{-0.3cm} 
% \end{figure*}
% \begin{figure}[htp] 
% \centering
% \subfigure[The effect of hop numbers]{
% \label{fig:res-para-h}
% \includegraphics[scale=0.21]{figures/layer.pdf}} 
% \quad
% \subfigure[The effect of bucket numbers]{
% \label{fig:res-para-b}
% \includegraphics[scale=0.21]{figures/bucket.pdf}} 
% \caption{The parameter sensitivity study of hop and bucket numbers.}
% \label{fig:res-para-hb}
% \end{figure}

Table~\ref{tab:res-few} shows $SR@K$, $P@K$ and $R@K$ ($K = \{1,10\}$) for all approaches on three datasets. To sum up, that \tool greatly outperforms other approaches in all metrics. In detail, $SR@10$ is improved by 298.7\%, $P@10$ is improved by 233.3\%, and $R@10$ is improved by 402.1\% compared to the latest approach GAPI on $SH_L$. Looking into the performance of baselines, GAPI obtains better performance than FOCUS, indicating the effectiveness of incorporating complicated connectivity information in enriching the representation of APIs with fewer direct interactions. 
Although \tool presents
% has 
a significant improvement on the recommendation performance,
% recommendation effect, 
the results of low-frequency APIs are still quite limited, which may be attributed to the functional particularity of the APIs and needs more future research.
% Thus, due to the functional particularity and unclear usage pattern of low-frequency APIs, it is still challenging to improve the recommendation accuracy of these APIs.
% This verifies significance of \tool in mining uncommon API usage pattern.
\begin{tcolorbox}[breakable,width=\linewidth-2pt,boxrule=0pt,top=4pt, bottom=4pt, left=7pt,right=7pt, colback=gray!20,colframe=gray!20]
\textbf{Answer to RQ3:} \tool consistently
% significantly 
outperforms the state-of-the-art baselines for recommending low-frequency APIs on the three benchmark datasets. Despite the superior performance of \tool, low-frequency recommendation is still challenging and needs more future research.
% ' recommendation, and \tool achieves stable performance on the three datasets. 
% \tool alleviates the problem of establishing the optimal representation of low-frequency APIs, by supplementing the poorly original representation with additional information. Therefore, the recommendation effectiveness for these APIs is promoted.
\end{tcolorbox}

\pgfplotstableread[row sep=\\,col sep=&]{
	datasets  & 8 & 16 & 32  & 64  \\
    8 & 0.730 & 0.735 & 0.709 & 0.735   \\
    24 & 0.767 & 0.794 & 0.773 & 0.757  \\
    40 & 0.741 & 0.762 & 0.751 & 0.751 \\
    56 & 0.730 & 0.714 & 0.741 & 0.735 \\
}\datasetsone

\pgfplotstableread[row sep=\\,col sep=&]{
	datasets & 8 & 16 & 32  & 64  \\
    8 &0.608 &0.613 &0.611 &0.608  \\
    24 &0.636 &0.634 & 0.641 &0.640  \\
    40 &0.639 &0.638 & 0.641 &0.641 \\
    56 &0.629 &0.636 &0.636 &0.641 \\
}\datasetstwo

\pgfplotstableread[row sep=\\,col sep=&]{
	datasets  & 8 & 16 & 32  & 64  \\
    8 & 0.824 & 0.822 & 0.829 & 0.810   \\
    24 & 0.840 & 0.840 & 0.835 & 0.822  \\
    40 & 0.842 & 0.836 & 0.824 & 0.834 \\
    56 & 0.842 & 0.838 & 0.830 & 0.830 \\
}\datasetsthree

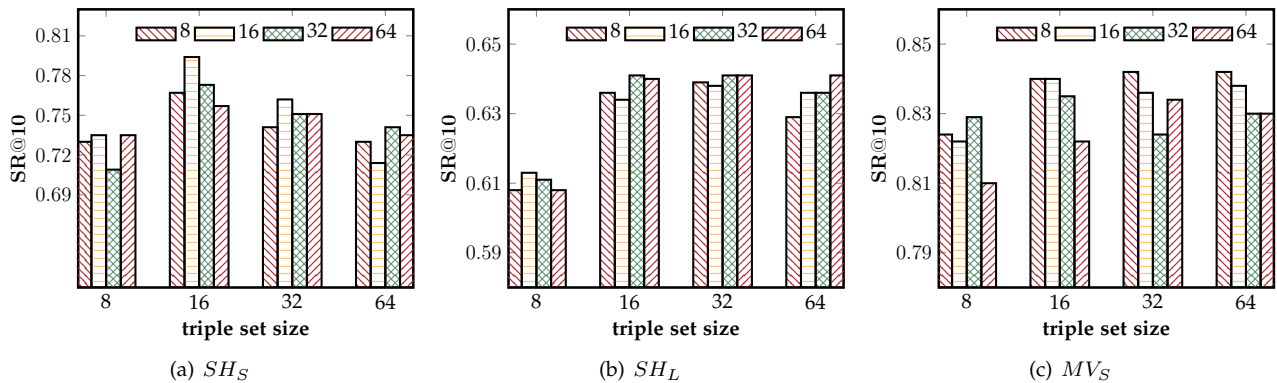
\begin{figure*}[h]	

	\subfigure[$SH_S$]{
    	\begin{tikzpicture}[scale=0.65]
    		\begin{axis}[
    	    	ybar=0pt,
    			bar width=0.3cm,
				% width=0.35\textwidth,
    % 			height=0.4\textwidth,
    			xlabel={\large \bf triple set size}, 
    			xtick={8,24,40,56},	
    			xticklabels={\large 8,\large 16,\large 32,\large 64},
                % legend style={at={(0.5,0.98)},
                % anchor=north,legend columns=-1,
                % draw=none},
                legend style = {
				    legend columns=-1,
				    draw=none,
				},
				legend image code/.code={
                    \draw [#1] (0cm,-0.18cm) rectangle (0.8cm,0.08cm); },
                % legend image code/.code={
                %     \draw [#1] (0cm,-0.28cm) rectangle (1.2cm,0.2cm); },
                % legend image code/.code={
                %     \draw [#1] (0cm,-0.263cm) rectangle (1.2cm,0.085cm); },
    % 			xmin=0.5,xmax=9.5,
    			ytick = {0.69, 0.72, 0.75, 0.78, 0.81},
    			ymin=0.62,ymax=0.83,
    			tick align=inside,
    			ticklabel style={font=\large},
   			    every axis plot/.append style={line width = 1.2pt},
    			every axis/.append style={line width = 1.5pt},
    			ylabel={\textbf{\large $\mathbf{SR@10}$}},
    			]
    			\addplot[pattern=north west lines, pattern color=c1] table[x=datasets,y=8]{\datasetsone};
    			\addplot[pattern = horizontal lines ,pattern color=c2] table[x=datasets,y=16]{\datasetsone};
    			\addplot[pattern = crosshatch,pattern color=c3] table[x=datasets,y=32]{\datasetsone};
    			\addplot[pattern = north east lines,pattern color=c1] table[x=datasets,y=64]{\datasetsone};
    			\legend{\large 8, \large 16,\large 32,\large 64}
    
    		\end{axis}
    	\end{tikzpicture}
    	\label{fig:res-para-shs}
	}
	\subfigure[$SH_L$]{
    	\begin{tikzpicture}[scale=0.65]
    		\begin{axis}[
    	    	ybar=0pt,
    			bar width=0.3cm,
				% width=0.53\textwidth,
    % 			height=0.4\textwidth,
    			xlabel={\large \bf triple set size},
    			xtick={8,24,40,56},	
    			xticklabels={\large 8,\large 16,\large 32,\large 64},
                % legend style={at={(0.5,0.98)},
                % anchor=north,legend columns=-1,
                % draw=none},
                legend style = {
				    legend columns=-1,
				    draw=none,
				},
				legend image code/.code={
                    \draw [#1] (0cm,-0.18cm) rectangle (0.8cm,0.08cm); },
                % legend image code/.code={
                %     \draw [#1] (0cm,-0.28cm) rectangle (1.2cm,0.2cm); },
                % legend image code/.code={
                %     \draw [#1] (0cm,-0.263cm) rectangle (1.2cm,0.085cm); },
    % 			xmin=0.5,xmax=9.5,
    			ytick = {0.59, 0.61, 0.63, 0.65},
    			ymin=0.58,ymax=0.66,
    			tick align=inside,
    			ticklabel style={font=\large},
   			    every axis plot/.append style={line width = 1.2pt},
    			every axis/.append style={line width = 1.5pt},
    			ylabel={\textbf{\large $\mathbf{SR@10}$}},
    			]
    			\addplot[pattern=north west lines, pattern color=c1] table[x=datasets,y=8]{\datasetstwo};
    			\addplot[pattern = horizontal lines ,pattern color=c2] table[x=datasets,y=16]{\datasetstwo};
    			\addplot[pattern = crosshatch,pattern color=c3] table[x=datasets,y=32]{\datasetstwo};
    			\addplot[pattern = north east lines,pattern color=c1] table[x=datasets,y=64]{\datasetstwo};
    			\legend{\large 8, \large 16,\large 32,\large 64}
    
    		\end{axis}
    	\end{tikzpicture}
    	\label{fig:res-para-shl}
	}
	\subfigure[$MV_S$]{
    	\begin{tikzpicture}[scale=0.65]
    		\begin{axis}[
    	    	ybar=0pt,
    			bar width=0.3cm,
				% width=0.53\textwidth,
    % 			height=0.4\textwidth,
    			xlabel={\large \bf triple set size}, 
    			xtick={8,24,40,56},	
    			xticklabels={\large 8,\large 16,\large 32,\large 64},
                % legend style={at={(0.5,0.98)},
                % anchor=north,legend columns=-1,
                % draw=none},
                legend style = {
				    legend columns=-1,
				    draw=none,
				},
				legend image code/.code={
                    \draw [#1] (0cm,-0.18cm) rectangle (0.8cm,0.08cm); },
                % legend image code/.code={
                %     \draw [#1] (0cm,-0.28cm) rectangle (1.2cm,0.2cm); },
                % legend image code/.code={
                %     \draw [#1] (0cm,-0.263cm) rectangle (1.2cm,0.085cm); },
    % 			xmin=0.5,xmax=9.5,
    			ytick = {0.79, 0.81, 0.83, 0.85},
    			ymin=0.78,ymax=0.86,
    			tick align=inside,
    			ticklabel style={font=\large},
   			    every axis plot/.append style={line width = 1.2pt},
    			every axis/.append style={line width = 1.5pt},
    			ylabel={\textbf{\large $\mathbf{SR@10}$}},
    			]
    			\addplot[pattern=north west lines, pattern color=c1] table[x=datasets,y=8]{\datasetsthree};
    			\addplot[pattern = horizontal lines ,pattern color=c2] table[x=datasets,y=16]{\datasetsthree};
    			\addplot[pattern = crosshatch,pattern color=c3] table[x=datasets,y=32]{\datasetsthree};
    			\addplot[pattern = north east lines,pattern color=c1] table[x=datasets,y=64]{\datasetsthree};
    			\legend{\large 8, \large 16,\large 32,\large 64}
    
    		\end{axis}
    	\end{tikzpicture}
    	\label{fig:res-para-mv}
	}
	\caption{The results of $SR@10$ on three datasets along with different
sizes of triple set.}

\end{figure*}

\pgfplotstableread[row sep=\\,col sep=&]{
	k & $SH_S$ & $SH_L$ & $MV_S$   \\
	1 & 0.794 & 0.633 & 0.840  \\
	2 & 0.720 & 0.559 & 0.790 \\
	3 & 0.656 & 0.548 & 0.769 \\
	4 & 0.614 & 0.498 & 0.768 \\
}\HOP

\pgfplotstableread[row sep=\\,col sep=&]{
    k & $SH_S$ & $SH_L$ & $MV_S$    \\
	1 & 0.767 & 0.631 & 0.833   \\
	5 & 0.773 & 0.631 & 0.836 \\
	10 & 0.751 & 0.633 & 0.835 \\
	15 &0.794 & 0.631 & 0.840  \\
	20 &0.773 & 0.601 & 0.834  \\
}\BUKET

\definecolor{c1}{RGB}{0,124,177} % SHS
\definecolor{c2}{RGB}{255,121,38} % SHL
\definecolor{c3}{RGB}{0,158,63} % MVS

\begin{figure}[h]
	\centering		
    % \ref{named1} \ref{named2} \ref{named3}\\
	\subfigure[The effect of hop numbers]{
		\begin{tikzpicture}[scale=0.5]
		    
			\begin{axis}[
			    legend style = {
				    legend columns=-1,
				    draw=none,
				},
				width=0.46\textwidth,
    			height=0.36\textwidth,
				% legend to name,
				xtick = {1,2,3,4},
				ymin=0.4,ymax=0.93,
				ytick = {0.5, 0.6, 0.7, 0.8,0.9},
				mark size=2.5pt, 
				ylabel={\Large $\mathbf{SR@10}$},
				xlabel={\Large \bf hop number}, 
				% ticklabel style={font=\Huge},
				ticklabel style={font=\Large},
				every axis plot/.append style={line width = 2pt},
				every axis/.append style={line width = 1.5pt},
				]
				\addplot [mark=diamond*,color=c1] table[x=k,y=$SH_S$]{\HOP};
				\addplot [mark=triangle*,color=c2] table[x=k,y=$SH_L$]{\HOP};
				\addplot [mark=pentagon*,color=c3] table[x=k,y=$MV_S$]{\HOP};
				
				\legend{$SH_S$, $SH_L$, $MV_S$}
			\end{axis}
		\end{tikzpicture}
		\label{fig:res-para-h}
	}
	\subfigure[The effect of bucket numbers]{
		\begin{tikzpicture}[scale=0.5]
			\begin{axis}[
			   legend style = {
				    legend columns=-1,
				    draw=none,
				},
				ymin=0.5,ymax=0.92,
				ytick = {0.6, 0.7, 0.8, 0.9},
				xtick = {1,5,10,15,20},
				mark size=2.5pt, 
				width=0.46\textwidth,
    			height=0.36\textwidth,
				ylabel={\Large \bf $\mathbf{SR@10}$},
				xlabel={\Large \bf bucket number}, 
				ticklabel style={font=\Large},
				every axis plot/.append style={line width = 2pt},
				every axis/.append style={line width = 1.5pt},
				]
				\addplot [mark=diamond*,color=c1] table[x=k,y=$SH_S$]{\BUKET};
				\addplot [mark=triangle*,color=c2] table[x=k,y=$SH_L$]{\BUKET};
				\addplot [mark=pentagon*,color=c3] table[x=k,y=$MV_S$]{\BUKET};
				\legend{$SH_S$, $SH_L$, $MV_S$}
			\end{axis}
		\end{tikzpicture}
		\label{fig:res-para-b}
	}

	\caption{The parameter sensitivity study of hop and bucket numbers.}
	\label{fig:res-para-hb}
% 	\vspace{-0.3cm} 
\end{figure}
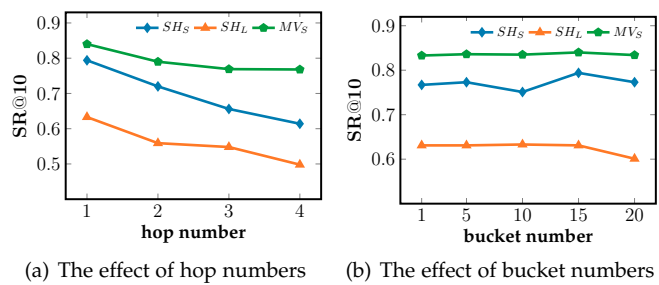

\subsection{Parameter Sensitivity Study (RQ4)}

We conduct experiments to analyze the impact of following hyper-parameters with different settings on \tool's performance. 

\textbf{Impact of max hop number $L$.} We vary the number of hops in propagating to observe the performance change of \tool. Figure~\ref{fig:res-para-h} depicts the results in terms of $SR@10$. We observe that \tool achieves the best results with one hop on three datasets, and the performance gradually decreases with the hop number increases.

One possible explanation is that, in the graph, short-distance nodes have a strong correlation with the original node, while the relevance decays as the distance increases. Consequently, the positive impact of short-distance propagation is greater, while long-distance propagation brings more noise than useful signals.

% \begin{figure*}[h] 
%   \centering
%   \includegraphics[scale=0.53]{figures/res-size.pdf}
%   \caption{The parameter sensitivity study of size of triple set on three datasets.}
%   \label{fig:res-para-size}
% \end{figure*}

% \begin{figure}[h] 
%   \centering
%   \includegraphics[scale=0.53]{figures/res-size-1.pdf}
%   \caption{The parameter sensitivity study of size of triple set on three datasets.}
%   \label{fig:res-para-size}
% \end{figure}

\textbf{Impact of bucket number $|\mathcal{T}|$.} To study the impact of bucket numbers, we conduct different experiments by setting different bucket numbers. The experimental results in terms of $SR@10$ are presented in Figure~\ref{fig:res-para-b}, which shows that for $SH_S$, $SH_L$, and $MV$, the best performance is achieved when the number of buckets is 15, 10, and 15, respectively.

One possible reason for this phenomenon is that when the number of buckets is too small, i.e., few relation types, the graph contains less information, which compromises the trained model's expressiveness. While a large number of buckets, i.e., many relation types, makes the information in the graph too rich, leading to over-fitting the model.

% \begin{figure}[htp]
% \centering
% \subfigure[$SH_S$]{
% \includegraphics[scale=0.5]{figures/res-shs-size.pdf}
% \label{fig:res-para-shs}} 

% \subfigure[$SH_L$]{
% \includegraphics[scale=0.5]{figures/res-shl-size.pdf}
% \label{fig:res-para-shl}} 

% \subfigure[$MV$]{
% \includegraphics[scale=0.5]{figures/res-mv-size.pdf}
% \label{fig:res-para-mv}}

% \caption{The result of $SR@10$ on three datasets with different sizes of triple set.}
% \end{figure}

\textbf{Impact of triple set size $|\mathcal{S}_u^l|$ in each hop $l$.} We change the number of neighbors selected by the client method and the target API in each hop to explore the effects of triple set size on \tool's performance. The results of $SR@10$ on the $SH_S$, $SH_L$ and $MV$ are demonstrated in Figure~\ref{fig:res-para-shs}, Figure~\ref{fig:res-para-shl} and Figure~\ref{fig:res-para-mv}, respectively. 

Jointly analyzing the three sub-figures, when the size increases, the results get better first and then worse. This means that when the size is moderately large, the benefits of more information included improves the performance. However, when the size is extremely large, the noise introduced outweighs the useful information introduced and thus it can hurt the performance. Overall, 16 or 32 is the suitable size of triple set in each hop for both the method and the API on three datasets.

\begin{tcolorbox}[breakable,width=\linewidth-2pt,boxrule=0pt,top=4pt, bottom=4pt, left=7pt,right=7pt, colback=gray!20,colframe=gray!20]
\textbf{Answer to RQ4:} The larger hop number has negative effect on \tool; the bucket number has stable effect on \tool; the triple set size has fluctuant effect on \tool.
\end{tcolorbox}

\section{Discussion}\label{sec:discussion}

% \subsection{Why our model work?}

%We analyze the following threats to validity of the model.

\subsection{Implications}
In this section, we discuss the implications that would be helpful for software researchers and software developers.

\textbf{Software Researchers.} 
% The analysis in the Section~\ref{sec:motivation} shows that the low-frequency API accounts for a large proportion, but the recommendation effect is very poor. \textcolor{red}{What do you mean here?}
In section~\ref{sec:results}, we achieve
% conclude 
that the
% rich 
heterogeneous information in source code is greatly beneficial for improving the recommendation performance of APIs including the low-frequency APIs. However, we also find that the results of low-frequency APIs are still quite limited, presenting a large gap with those of common APIs. The limited results may be attributed to the functional particularity of the low-frequency APIs, and could impact the practical usage of current API recommendation tools. Thus, we suggest researchers working on API recommendation to focus more on the recommendation of low-frequency APIs by combining external knowledge such as API documentation or exploring data augmentation techniques.

\textbf{Software Developers.} 
According to our coarse analysis of the benchmark datasets, low-frequency APIs are usually not associated with API documentation. API documentation which contains usage samples and instructions is helpful for learning the representations of APIs~\cite{DBLP:conf/kbse/ThungWLL13,DBLP:conf/kbse/HuangXXLW18}. Thus, we encourage developers to write some descriptions or usage examples for facilitating the API recommendation task.

\subsection{Threats to Validity}
\textbf{Internal Validity.} In this paper, following~\cite{DBLP:conf/cikm/WangZWZLXG18, DBLP:conf/sigir/WangLTCL20} we sample a fixed-size of neighbors on graphs instead of using a full size triple sets for the trade off of computation overhead. This may slightly influence the performance of \tool. To alleviate the impact of this threat, we conduct each experiment five times and obtain average performance as shown in Section~\ref{sec:results}. Furthermore, our experiments on parameter sensitivity also demonstrates that different sizes of triple set influence the performance of \tool slightly.

\textbf{External Validity.} We evaluate \tool under Java datasets, while \tool may show different performance
% perform differently 
on datasets in
% built upon 
other programming languages. To reduce the impact from different programming languages, when designing three multi-view graphs, we try to exclude the language-specific information and only maintain the structure information such as call relationships and definition relationships. We believe our design can be easily adapted to most programming languages.

\section{Related Work}\label{sec:related}

In this section, we review existing work about API usage recommendation.
% \subsection{API Usage Recommendation}
The work on API usage recommendation can be divided into two categories: pattern-based methods and learning-based methods. Pattern-based methods utilize traditional statistical methods to capture usage patterns from API co-occurrences. Learning-based methods leverage deep learning models to automatically learn the potential usage patterns from a large code corpus and then use them to recommend patterns.

\textbf{Pattern-based methods.}
Zhong \etal propose MAPO~\cite{DBLP:conf/ecoop/ZhongXZPM09} to cluster and mine API usage patterns from open source repositories, and then recommends the relevant usage patterns to developers. 
Wang \etal improve MAPO and build UP-Miner~\cite{DBLP:conf/msr/WangDZCXZ13} by utilizing a new algorithm based on $SeqSim$ to cluster the API sequences.
Nguyen \etal propose APIREC~\cite{DBLP:conf/sigsoft/NguyenHCNMRND16}, which uses fine-grained code changes and the corresponding changing contexts to recommend APIs.
Fowkes \etal propose PAM~\cite{DBLP:conf/sigsoft/FowkesS16} to tackle the problem that the recommended API lists are large and hard to understand. PAM mines API usage patterns through an almost parameter-free probabilistic algorithm and uses them to recommend APIs.
Liu \etal propose RecRank~\cite{DBLP:conf/kbse/LiuHN18} to improve the top-1 accuracy based on API usage paths.
Nguyen \etal propose FOCUS~\cite{DBLP:conf/icse/NguyenRRODP19}, which mines open-source repositories and analyzes API usages in similar projects to recommend APIs and API usage patterns based on context-aware collaborative-filtering techniques.
Previous pattern-based methods only consider one or two relationships between APIs, however, \tool considers call interactions, API co-occurrences, project/package hierarchical structure to comprehensively capture the contexts surrounding APIs.

\textbf{Learning-based methods.}
Nguyen \etal propose a graph-based language model GraLan~\cite{DBLP:conf/icse/NguyenN15} to recommend API usages.
Gu \etal propose DeepAPI~\cite{DBLP:conf/sigsoft/GuZZK16}. They reformulate API recommendation task as a query-API translation problem and use an RNN Encoder-Decoder model to recommend API sequences.
Ling \etal propose GeAPI~\cite{DBLP:journals/jcst/LingZLX19}. GeAPI automatically constructs API graphs based on source code and leverages graph embedding techniques for API representation.
Gu \etal propose Codekernel~\cite{DBLP:conf/kbse/Gu0019} by representing code as object usage graphs and clustering them to recommend API usage examples.
Zhou \etal build a tool named BRAID~\cite{DBLP:conf/sigsoft/ZhouJYCNG21} to leverage learning-to-rank and active learning techniques to boost recommendation performance.
Previous learning-based methods fail to recommend usage patterns for low-frequency APIs due to the data-driven feature, \tool encodes API frequency with global API co-occurrence graph to alleviate this problem.

\section{Conclusion}\label{sec:conclusion}
In this paper, we propose a novel approach \tool for automatic API usage recommendation. \tool employs heterogeneous graphs, which are constructed from multiple views, i.e., method-API interaction from local view, API-API co-occurrence from global view, and project structure from external view. A graph representation model with a frequency-aware attentive network and
a structure-aware attentive network is then proposed to learn the matching scores between methods and APIs based on the multi-view graphs. Experiment demonstrates \tool's effectiveness both on overall API usage recommendation and low-frequency API usage recommendation. For future work, in addition to the information extracted from projects, some information from API official documentation or Q\&A sites also contributes to mining API usage patterns. Therefore, we plan to design some new modules that encode more information from different sources. 

\bibliographystyle{IEEEtran}
\bibliography{ref}

\end{document}